\documentclass[aps,prl,superscriptaddress,showpacs,floatfix,nofootinbib,notitlepage,twocolumn,longbibliography]{revtex4-1}
\usepackage{amsmath,graphicx,float,csquotes,mdframed,appendix,url}
\usepackage[dvipsnames]{xcolor}
\usepackage[colorlinks=true, pdfstartview=FitV, linkcolor=RedOrange, citecolor=Emerald, urlcolor=Cerulean]{hyperref}

\newcommand{\snn}{\sqrt{s_\mathrm{NN}}}
\newcommand{\fmc}{fm/$c$}
\newcommand{\Glb}{\textsc{3d-Glauber}}
\newcommand{\MUSIC}{\textsc{music}}
\newcommand{\URQMD}{\textsc{urqmd}}

\begin{document}

\title{Viscosities of the Baryon-Rich Quark-Gluon Plasma from Beam Energy Scan Data}

\author{Chun Shen}
\affiliation{Department of Physics and Astronomy, Wayne State University, Detroit, Michigan 48201, USA}
\affiliation{RIKEN BNL Research Center, Brookhaven National Laboratory, Upton, NY 11973, USA}

\author{Bj\"orn Schenke}
\affiliation{Physics Department, Brookhaven National Laboratory, Upton, NY 11973, USA}

\author{Wenbin Zhao}
\affiliation{Department of Physics and Astronomy, Wayne State University, Detroit, Michigan 48201, USA}
\affiliation{Nuclear Science Division, Lawrence Berkeley National Laboratory, Berkeley, California 94720, USA}
\affiliation{Physics Department, University of California, Berkeley, California 94720, USA}
\begin{abstract}

This work presents the first Bayesian inference study of the (3+1)D dynamics of relativistic heavy-ion collisions and Quark-Gluon Plasma (QGP) viscosities using an event-by-event (3+1)D hydrodynamics + hadronic transport theoretical framework and data from the Relativistic Heavy Ion Collider (RHIC) Beam Energy Scan program. Robust constraints on initial state nuclear stopping and the baryon chemical potential-dependent shear viscosity of the produced quantum chromodynamic (QCD) matter are obtained. The specific bulk viscosity of the QCD matter is found to exhibit a preferred maximum around $\snn = 19.6$\,GeV. This result allows for the alternative interpretation of a reduction (and/or increase) of the speed of sound relative to that of the employed lattice-QCD based equation of state (EOS) for net baryon chemical potential $\mu_B \sim 0.2\, (0.4)$\,GeV.

\end{abstract}

\maketitle


\paragraph{1.~Introduction.}

The characterization of Quark-Gluon Plasma (QGP) has long been a central pursuit in high-energy nuclear physics~\cite{Achenbach:2023pba,Arslandok:2023utm}. The Relativistic Heavy Ion Collider (RHIC) at Brookhaven National Laboratory has played a pivotal role in this endeavor, providing opportunities for studying strongly interacting matter at extreme temperatures and densities. One of the most intriguing aspects of RHIC experiments is the Beam Energy Scan (BES) program~\cite{Caines:2009yu, Mohanty:2011nm, Mitchell:2012mx, Odyniec:2015iaa}, which systematically varies the center-of-mass energy of colliding ions to investigate the properties of the QGP over a wide range of the temperature and baryon chemical potential dependent phase diagram of Quantum Chromodynamics (QCD). The BES program allows us to investigate the transition between hadronic matter and the QGP and to search for a possible critical point and first-order phase boundaries, shedding light on the emergent properties of the nuclear force (see reviews \cite{STAR:2010vob,Luo:2017faz,Bzdak:2019pkr,An:2021wof}).

The theoretical description of the QGP and its real-time evolution in relativistic heavy-ion collisions is a complex and multifaceted challenge~\cite{Shen:2014vra, Putschke:2019yrg, Schenke:2020mbo, Nijs:2020roc, Pang:2018zzo}. While relativistic viscous hydrodynamics is an efficient and effective framework to describe the QGP collectivity, uncertainties in the initial conditions and the transport properties of the medium introduce significant ambiguities in the theoretical predictions.
Quantifying these uncertainties is essential for extracting precise information about the QGP's properties~\cite{Song:2010mg, Schenke:2010rr, Qiu:2011hf,  Gale:2013da, Heinz:2013th, Schenke:2019pmk, Shen:2020mgh}.

While it is challenging to compute the QGP transport coefficients from first principles (see \cite{Altenkort:2022yhb} for recent lattice extractions of viscosities for a purely gluonic system), phenomenological studies showed that hadronic observables measured in heavy-ion collisions are sensitive to the shear and bulk viscosity of QCD matter \cite{Song:2010mg,Ryu:2015vwa,Karpenko:2015xea,Shen:2015msa, Ryu:2017qzn, Schenke:2019ruo, Schenke:2020mbo}.
Early work constraining these transport coefficients with hydrodynamic simulations of heavy-ion collisions generally focused on the shear viscosity, approximated as an effective constant ratio to the entropy density $\eta/s$ \cite{Song:2010mg, Niemi:2015qia, Karpenko:2015xea,Shen:2020jwv}.
Contemporary efforts adopted the Bayesian Inference method to constrain the QGP's specific shear and bulk viscosities, including the uncertainties from all the other model parameters. Large-scale model-to-data comparisons are necessary to achieve this goal, given the significant computational challenge of constraining a high-dimensional model parameter space \cite{Pratt:2015zsa,Bernhard:2016tnd,Auvinen:2017fjw,Bernhard:2019bmu,Nijs:2020ors,JETSCAPE:2020shq,JETSCAPE:2020mzn,Parkkila:2021yha,Parkkila:2021tqq,Phillips:2020dmw,Heffernan:2023gye,Heffernan:2023utr}. 

Aiming to make extensive use of existing rapidity and collision energy dependent data, we perform comprehensive modeling of the (3+1)D QGP dynamics in a 26-dimensional model parameter space with state-of-the-art relativistic viscous hydrodynamics + hadronic transport simulations. This is a significant extension over an earlier work \cite{Auvinen:2017fjw}, which studied a much smaller 5-dimensional model parameter space. By performing the Bayesian inference analysis with multi-system measurements from the RHIC BES program phase I, we will obtain robust constraints on initial-state nuclear stopping and the temperature and baryon chemical potential dependent QGP shear and bulk viscosities for the first time. 

\paragraph{2. Hybrid framework and model parametrizations.}

To model the dynamics of Au+Au collisions from $\snn = 7.7$ to 200 GeV in the RHIC BES program, we employ a (3+1)D dynamical initialization model (\Glb) coupled with the hybrid framework of relativistic viscous hydrodynamics (\MUSIC) + hadronic transport (\URQMD)~\cite{Shen:2017bsr,Shen:2022oyg}. The \Glb{} model simulates the initial stage of heavy-ion collisions as the two nuclei pass through each other. Individual nucleon-nucleon (NN) collisions are determined based on their transverse positions and the inelastic NN cross section at the given collision energy. For each NN collision, we select valence quarks and soft partons inside the colliding nucleons to lose energy~\cite{Shen:2022oyg}. To constrain the initial-state nuclear stopping in this analysis, we parametrize the average amount of rapidity loss for each parton pair as a piece-wise function,
\begin{equation}
    \langle y_{\rm loss} \rangle = \left\{\begin{array}{cc}
        y_{\rm loss, 2} \frac{y_{\rm init}}{2} & 0 < y_{\rm init} \le 2 \\
        y_{\rm loss, 2} + (y_{\rm loss, 4} - y_{\rm loss, 2}) \frac{y_{\rm init} - 2}{2} & 2 < y_{\rm init} < 4 \\
        y_{\rm loss, 4} + (y_{\rm loss, 6} - y_{\rm loss, 4}) \frac{y_{\rm init} - 4}{2} & y_{\rm init} \ge 4 
    \end{array},\right. \notag
\end{equation}
where the parameter $y_{\rm loss, n}$ specifies the average amount of rapidity loss for $y_{\rm init} = n$. The event-by-event fluctuations of rapidity loss are introduced by the variance parameter $\sigma_{y_\mathrm{loss}}$ \cite{Shen:2022oyg}. After individual NN collision, wounded partons are decelerated with a string tension in the longitudinal direction during the time $\tau_\mathrm{hydro} = 0.5$\,\fmc{} in the collision rest frame. The lost energy and momentum produce an energy-momentum current $J^\mu$, which is fed into the hydrodynamic fields via a source term,
\begin{align}
    \partial_\mu T^{\mu\nu} &= J^\mu \\
    \partial_\mu J^{\mu}_B &= \rho_B.
\end{align}
In the second equation,  baryon charge densities from incoming nucleons are treated as scalar sources to the hydrodynamic net baryon current~\cite{Shen:2022oyg}.

We parametrize a blast-wave-like pre-equilibrium transverse flow profile for each string, developed during its hydrodynamization period $\tau_\mathrm{hydro}$ with the transverse flow rapidity~\cite{Zhao:2022ugy},
\begin{equation}
    \eta_\perp ({\bf x}_\perp) = \alpha_{\rm preFlow} | \tilde{\bf x}_\perp | ,
\end{equation}
where the 2D-vector $\tilde{\bf x}_\perp = (x - x_\mathrm{string}, y - y_\mathrm{string})$, with $x_\mathrm{string}$ and $y_\mathrm{string}$ the coordinates of the string in the transverse plane, and the parameter $\alpha_\mathrm{preFlow}$ controls the size of the pre-equilibrium flow. Then, the energy-momentum current $J^\mu$ can be written as
\begin{equation}
    J^\mu({\bf x}_\perp, \eta_s) = e_\mathrm{string}({\bf x}_\perp, \eta_s) u^\mu_\mathrm{string}(\eta_\perp({\bf x}_\perp), y(\eta_s)),
\end{equation}
where the string's local flow velocity is $u^\mu_\mathrm{string}(\eta_\perp, y) = (\cosh \eta_\perp \cosh y, \sinh \eta_\perp \hat{\bf e}_{\tilde{\bf x}_\perp}, \cosh \eta_\perp \sinh y)$ with $\hat{\bf e}_{\tilde{\bf x}_\perp} = \tilde{\bf x}_\perp/|\tilde{\bf x}_\perp|$ being the unit vector of $\tilde{\bf x}_\perp$ in the transverse plane. Precise definitions of $e_\mathrm{string}({\bf x}_\perp, \eta_s)$ and $y(\eta_s)$ can be found in Ref.~\cite{Shen:2022oyg}.
The hydrodynamic equations of motion are solved with a lattice-QCD-based equation of state (EOS) at finite densities, \textsc{neos-bqs}, which imposes strangeness neutrality and $n_Q = 0.4 n_B$ for Au+Au collisions~\cite{Monnai:2019hkn}.

To account for shear and bulk viscous effects in the hydrodynamic phase~\cite{Schenke:2010nt, Paquet:2015lta,Denicol:2018wdp}, we parametrize the baryon chemical potential $\mu_B$ dependence of the QGP shear viscosity as
\begin{equation}
    \label{eq:shearviscosity}
    \tilde{\eta}(\mu_B) = \left\{ \begin{array}{cc}
       \eta_0 + (\eta_2 - \eta_0)\frac{\mu_B}{0.2}  &  0 < \mu_B \le 0.2\,\mathrm{GeV} \\
       \eta_2 + (\eta_4 - \eta_2)\frac{(\mu_B - 0.2)}{0.2}  &  0.2 < \mu_B < 0.4\,\mathrm{GeV} \\
       \eta_4 &  \mu_B \ge 0.4\,\mathrm{GeV}
    \end{array},\right.
\end{equation}
where $\tilde{\eta} \equiv \eta T /(e + P)$ and the parameters $\eta_0$, $\eta_2$, $\eta_4$ are the values of the QGP specific shear viscosity at $\mu_B = 0, 0.2, 0.4$\,GeV, respectively. The translation from $\tilde{\eta}$ to $\eta/s$ introduces a mild temperature dependence at finite net baryon density, namely $\eta/s(T, \mu_B) = (1 + \frac{\mu_B n_B}{T s}) \tilde{\eta}(\mu_B)$. To limit the number of model parameters, we do not include an explicit temperature dependence for $\tilde{\eta}$ here, since the results from previous Bayesian analyses were compatible with a temperature independent $\eta/s$ value in the phase described by hydrodynamics~\cite{JETSCAPE:2020mzn}. 

The specific bulk viscosity is parametrized as an asymmetric Gaussian in temperature~\cite{Schenke:2019ruo, Schenke:2020mbo},
\begin{equation}
    \tilde{\zeta}(T, \mu_B) = \left\{ \begin{array}{cc}
       \zeta_{\rm max} \exp\left[-\frac{(T - T_\zeta(\mu_B))^2}{2\sigma_{\zeta, -}^2}\right] & T < T_\zeta(\mu_B) \\
       \zeta_{\rm max} \exp\left[-\frac{(T - T_\zeta(\mu_B))^2}{2\sigma_{\zeta, +}^2}\right] & T \ge T_\zeta(\mu_B) 
    \end{array},\right.
\end{equation}
where $\tilde{\zeta} \equiv \zeta T /(e + P)$ and the bulk peak temperature $T_{\zeta}(\mu_B) = T_{\zeta, 0} - \frac{0.15}{1\,\mathrm{GeV}} \mu_B^2$, so that it closely follows the constant energy density curve with $e = e(T_{\zeta, 0}, \mu_B = 0)$ for the \textsc{neos-bqs} EOS~\cite{Monnai:2019hkn}. This ensures that the bulk viscosity peak closely follows the phase crossover at finite net baryon density~\cite{HotQCD:2018pds,Borsanyi:2020fev,An:2021wof}.

Below the switching energy density $e_\mathrm{sw}$, individual fluid cells are converted into hadrons according to the Cooper-Frye particlization procedure, including out-of-equilibrium corrections to particle distributions with multiple conserved charge currents ($B, Q, S$) using the Grad moment method~\cite{Zhao:2022ugy}. The produced hadrons are then fed to the \URQMD{} transport model for hadronic scatterings and decays \cite{Bass:1998ca, Bleicher:1999xi}. The hadronic transport model controls the non-trivial $(T,\mu_B)$ dependence of viscosity in the dilute hadronic phase, which we do not vary in this Bayesian analysis.

\begin{table}[t!]
    \centering
    \caption{The 20 model parameters and their prior ranges.}
    \begin{tabular}{c|c|c|c} \hline \hline
        Parameter & Prior & Parameter & Prior \\ \hline
        $B_G$ (GeV$^{-2}$) & [1, 25] & $\alpha_\mathrm{string\,tilt}$ & [0, 1]  \\
        $\alpha_{\rm shadowing}$ & [0, 1] & $\alpha_\mathrm{preFlow}$ & [0, 2] \\
        $y_\mathrm{loss, 2}$ & [0, 2] & $\eta_0$ & [0.001, 0.3]  \\
        $y_\mathrm{loss, 4}$ & [1, 3] & $\eta_2$ & [0.001, 0.3] \\
        $y_\mathrm{loss, 6}$ & [1, 4] & $\eta_4$ & [0.001, 0.3] \\
        $\sigma_{y_\mathrm{loss}}$ & [0.1, 0.8] & $\zeta_{\rm max}$  & [0, 0.2] \\
        $\alpha_\mathrm{Rem}$ & [0, 1] & $T_{\zeta, 0}$ (GeV) & [0.15, 0.25] \\
        $\lambda_B$ & [0, 1] & $\sigma_{\zeta, +}$ (GeV) & [0.01, 0.15] \\
        $\sigma_x^\mathrm{string}$ (fm) & [0.1, 0.8] & $\sigma_{\zeta, -}$ (GeV) & [0.005, 0.1] \\
        $\sigma_\eta^\mathrm{string}$ & [0.1, 1] & $e_\mathrm{sw}$ (GeV/fm$^3$) & [0.15, 0.5] \\
        \hline \hline     
    \end{tabular}
    \label{table:paramPrior}
\end{table}

All model parameters are listed in Table~\ref{table:paramPrior} with their prior ranges. The definitions of the parameters $B_G$, $\alpha_{\rm shadowing}$, $\lambda_B$, $\sigma_x^{\rm string}$, $\sigma_\eta^{\rm string}$, $\alpha_{\rm string\,tilt}$ can be found in Ref.~\cite{Shen:2022oyg}.

To obtain an estimate of the $\mu_B$ dependence of the bulk viscosity, we allow the model parameters $\zeta_{\rm max}$ and $\sigma_{\zeta, \pm}$ to be independent parameters at different collision energies. This treatment enlarges the model parameter space from 20 dimensions to 26 dimensions.

\begin{table}[t]
 \centering
    \caption{The experimental measurements in Au+Au collisions used in this Bayesian inference study.}
    \begin{tabular}{c|c|c} \hline \hline
        $\snn$ (GeV) & STAR & PHOBOS \\ \hline
             & $dN/dy (\pi^+, K^+, p, \bar{p})$ \cite{STAR:2008med}& $dN^\mathrm{ch}/d\eta$ \cite{PHOBOS:2005zhy}\\
        200  & $\langle p_T \rangle (\pi^+, K^+, p, \bar{p})$ \cite{STAR:2008med} & $v^\mathrm{ch}_2(\eta)$ \cite{PHOBOS:2006dbo} \\
             & $v^\mathrm{ch}_2\{2\}$~\cite{STAR:2017idk}, $v^\mathrm{ch}_3\{2\}$~\cite{STAR:2016vqt}  & \\ \hline
             & $dN/dy (\pi^+, K^+, p)$ \cite{STAR:2017sal}&  \\
        19.6  & $\langle p_T \rangle (\pi^+, K^+, p, \bar{p})$ \cite{STAR:2017sal}& $dN^\mathrm{ch}/d\eta$ \cite{PHOBOS:2005zhy} \\
             & $v^\mathrm{ch}_2\{2\}$~\cite{STAR:2017idk}, $v^\mathrm{ch}_3\{2\}$~\cite{STAR:2016vqt}  & \\ \hline
             & $dN/dy (\pi^+, K^+, p)$ \cite{STAR:2017sal}&  \\ 
        7.7  & $\langle p_T \rangle (\pi^+, K^+, p, \bar{p})$ \cite{STAR:2017sal}& \\
             & $v^\mathrm{ch}_2\{2\}$~\cite{STAR:2017idk}, $v^\mathrm{ch}_3\{2\}$~\cite{STAR:2016vqt} & \\
        \hline \hline     
    \end{tabular}
    \label{table:expData}
\end{table}

Table~\ref{table:expData} summarizes the experimental observables (604 data points in total) in the current Bayesian inference study. The mid-rapidity measurements in Au+Au collisions at 200, 19.6, and 7.7 GeV can cover up to $\mu_B$\,$\sim$\,$0.4$\,GeV in the QCD phase diagram \cite{Cleymans:2005xv,Andronic:2009jd,STAR:2017sal}. Because the theoretical uncertainty is significant in peripheral collisions, we use identified particle yields and their mean $p_T$ from 0-5\% to 50-60\% centrality and charged hadron $v_n\{2\}$ from central up to 40-50\% centrality. We do not include the antiproton yields at 19.6 and 7.7 GeV because the statistical errors in the training simulations are still too big for reliable model emulation. In our following analysis, we will quantify the impacts of including the pseudorapidity distribution of charged hadron yields and their elliptic flow coefficient from the PHOBOS Collaboration on constraining the QGP properties. 

To efficiently explore the parameter space $\{\theta \}$ listed in Table~\ref{table:paramPrior}, we train Gaussian Process (GP) emulators for our model calculations with 1,000 design points in the model parameter space. These 1,000 design points are sampled using the Maximum Projection Latin Hypercube Design algorithm~\cite{MaxProLHD1, MaxProLHD2}. At every design parameter point, we simulate 1,000 minimum bias Au+Au collisions at 200 GeV and 2,000 minimum bias events at 19.6 and 7.7 GeV each. An interactive web page with the trained GP emulators is available to help the interested reader develop intuition about how the model parameters affect the observables~\cite{streamlit}.

Using the trained GP emulators, we can obtain the posterior distribution of model parameters, $\mathcal{P}(\theta \vert y_\mathrm{exp})$, following Bayes' theorem by sampling the uniform prior $\mathcal{P}(\theta)$ with the Monte Carlo Markov Chain (MCMC) method,
\begin{equation}
    \mathcal{P}(\theta \vert y_\mathrm{exp}) \propto \mathcal{P}(y_\mathrm{exp} \vert \theta) \mathcal{P}(\theta).
\end{equation}
Here $\mathcal{P}(y_\mathrm{exp} \vert \theta)$ is the likelihood for the model results with parameter $\theta$ to agree with the experimental data $y_\mathrm{exp}$. It is defined as a multivariate normal distribution~\cite{Mantysaari:2022ffw}. We verify our Bayesian inference analysis with a closure test in the supplemental material.

\paragraph{3. Results and Discussions.}

After performing the Bayesian inference analysis on the STAR and PHOBOS data listed in Table~\ref{table:expData}, we obtain the posterior distribution for our model parameters. In this letter, we will focus on the constraints on initial-state nuclear stopping and QGP shear and bulk viscosities, which are of primary physics interest. A complete analysis will be reported in the follow-up work.

\begin{figure}[t]
    \centering
    \includegraphics[width=\linewidth]{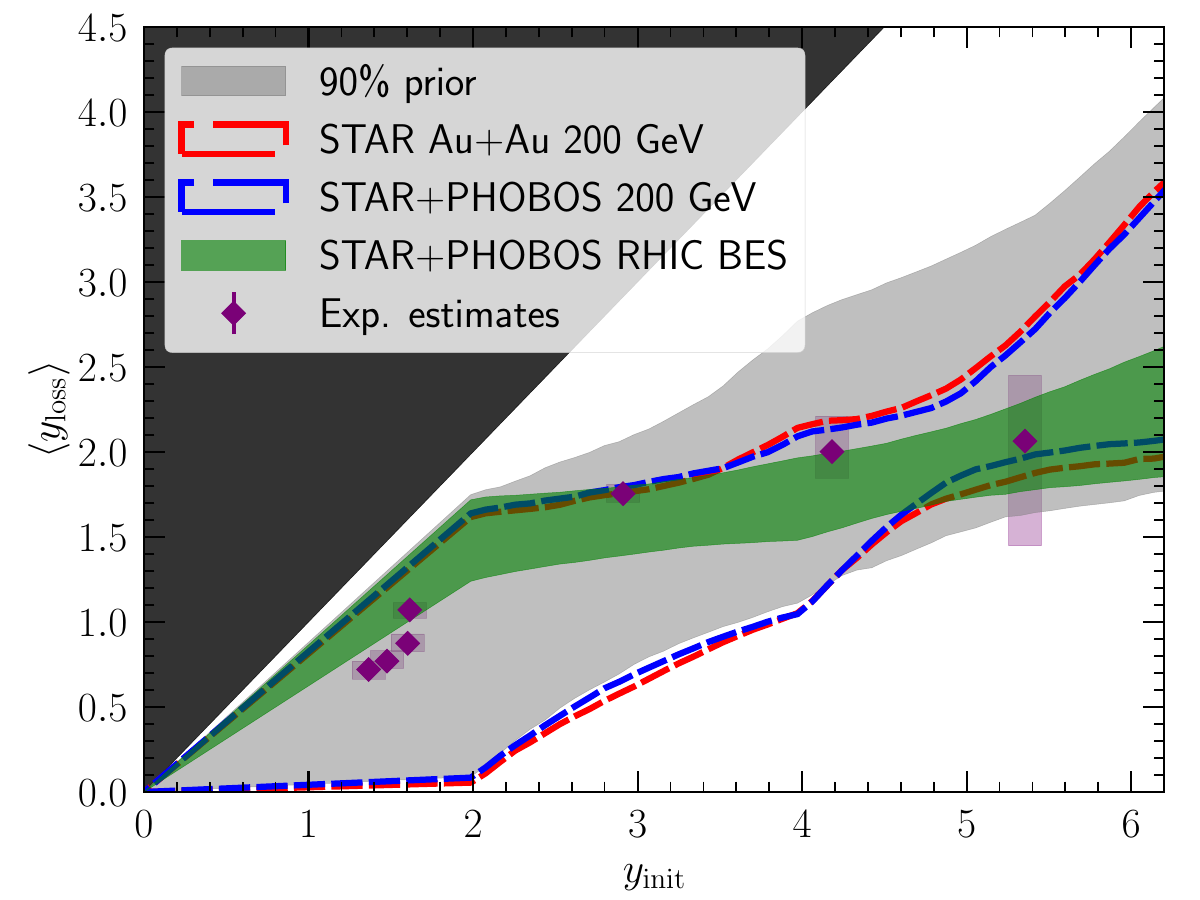}
    \caption{(Color online) Posterior distributions of the average initial-state rapidity loss at the nuclear impact. Color bands indicate 90\% confidence intervals. The experimental estimate of initial-state nuclear stopping is taken from the net proton rapidity measurements~\cite{PhysRevC.52.2684, BRAHMS:2009wlg}.}
    \label{fig:ylossPost}
\end{figure}

Figure~\ref{fig:ylossPost} shows prior and posterior distributions of the average rapidity loss as a function of initial-state rapidity $y_{\rm init}$ in the \Glb{} model. The narrowing in the 90\% prior for $y_{\rm init}$ between the transition points of the linear parametrization is an artifact of this choice of parametrization. The average rapidity loss is strongly correlated with the amount of particle production in the collisions. The comparison of the 90\% prior (the light gray band) with the red band shows that the identified particle yields at the top RHIC energy can constrain the $\langle y_\mathrm{loss} \rangle$ for $y_\mathrm{init} \in [4, 6]$. This result is consistent with the fact that the incoming nucleons' beam rapidity $y_\mathrm{beam} \equiv {\rm arccosh}(\snn/(2 m_{\rm N}))=5.36$ at $\sqrt{s_\mathrm{NN}} = 200$ GeV~\cite{Shen:2018pty}. 

Our analysis suggests that the average rapidity loss at $\sqrt{s_\mathrm{NN}} = 200$ GeV is $\langle y_\mathrm{loss} \rangle \sim 2$, which is consistent with estimations based on BRAHMS measurements~\cite{BRAHMS:2009wlg}. The mild difference between the red and blue bands in Fig.~\ref{fig:ylossPost} indicates that the PHOBOS $dN^{\rm ch}/d\eta$ measurements do not impose any significant additional constraints on the $\langle y_\mathrm{loss} \rangle$ parameter, because the data have relatively large error bars compared to the STAR measurements at mid-rapidity. 

Employing the RHIC BES data in the Bayesian analysis results in the green band, which is significantly narrower than the others. This result demonstrates that particle yield measurements from 7.7 to 200 GeV can impose strong constraints on the average rapidity loss for $y_\mathrm{init} \le 6$. Our constraints also agree well with independent experimental estimates from baryon stopping measurements~\cite{PhysRevC.52.2684, BRAHMS:2009wlg}. For low energy collisions with $y_\mathrm{init} < 2$, our current constraint is slightly larger than the experimental estimates from the E917 and E802/E866 experiments~\cite{PhysRevC.52.2684}. Future calibrations including these measurements will further refine the rapidity loss constraints at small $y_\mathrm{init}$.

\begin{figure}[t]
    \centering
    \includegraphics[width=\linewidth]{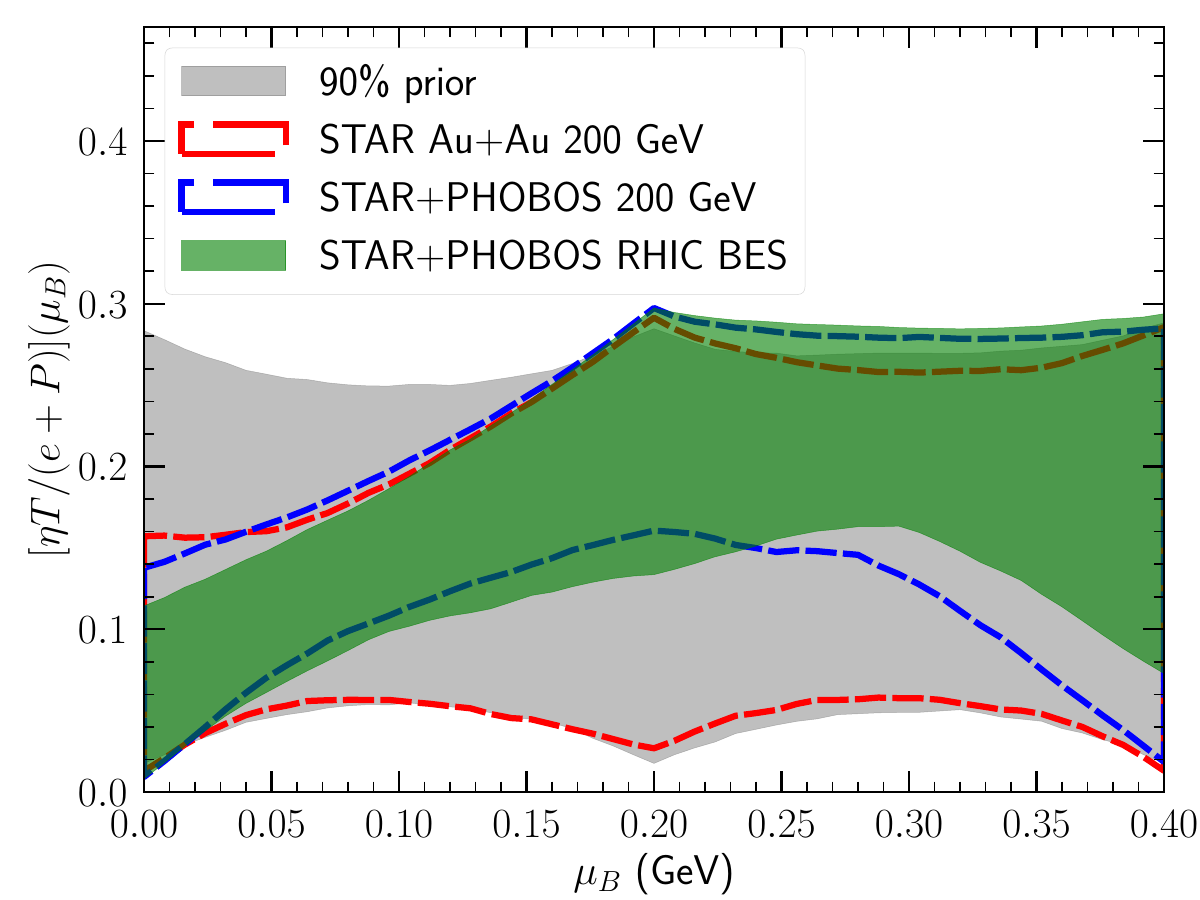}
    \caption{(Color online) Posterior distribution of the $\mu_B$ dependent QGP specific shear viscosity. Bands indicate 90\% confidence intervals. }
    \label{fig:shearPost}
\end{figure}

\begin{figure}[ht!]
    \centering
    \includegraphics[width=\linewidth]{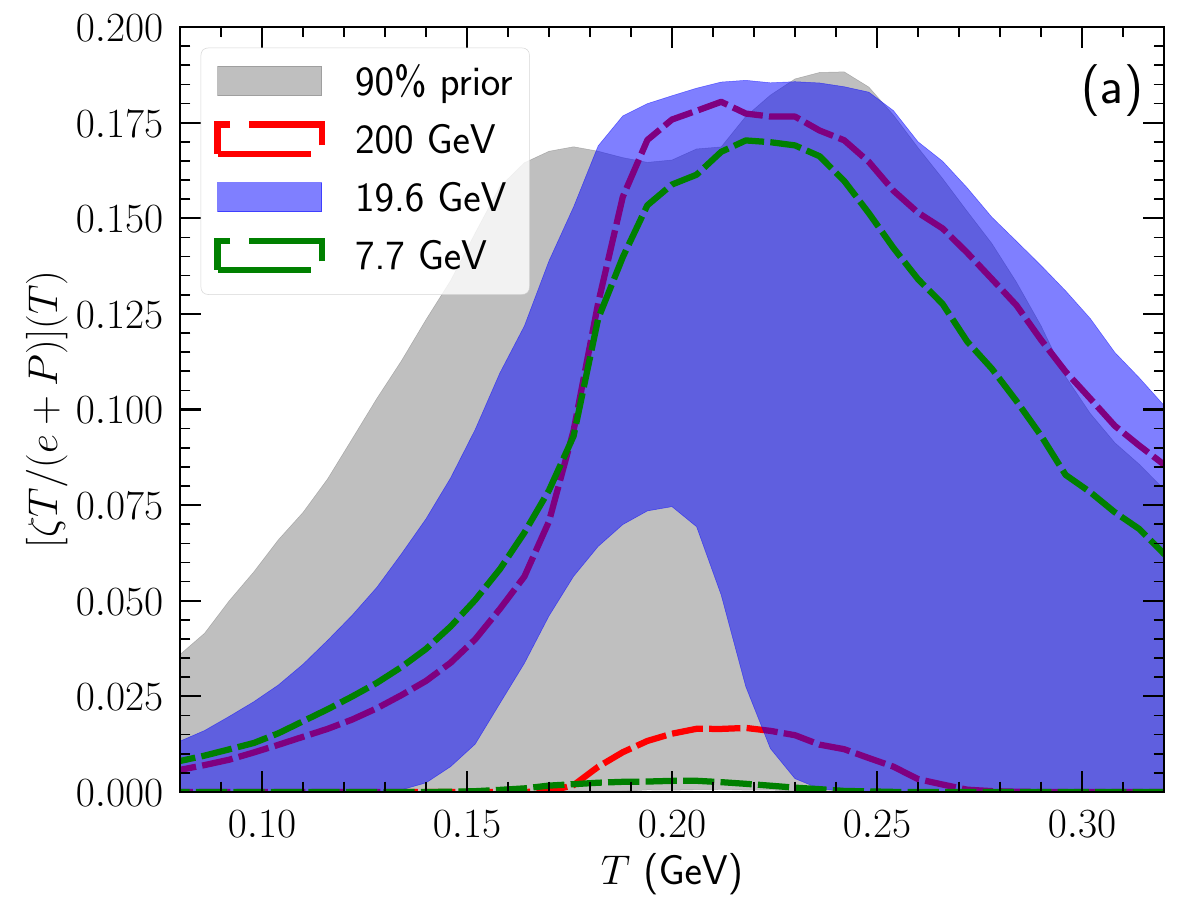}
    \includegraphics[width=\linewidth]{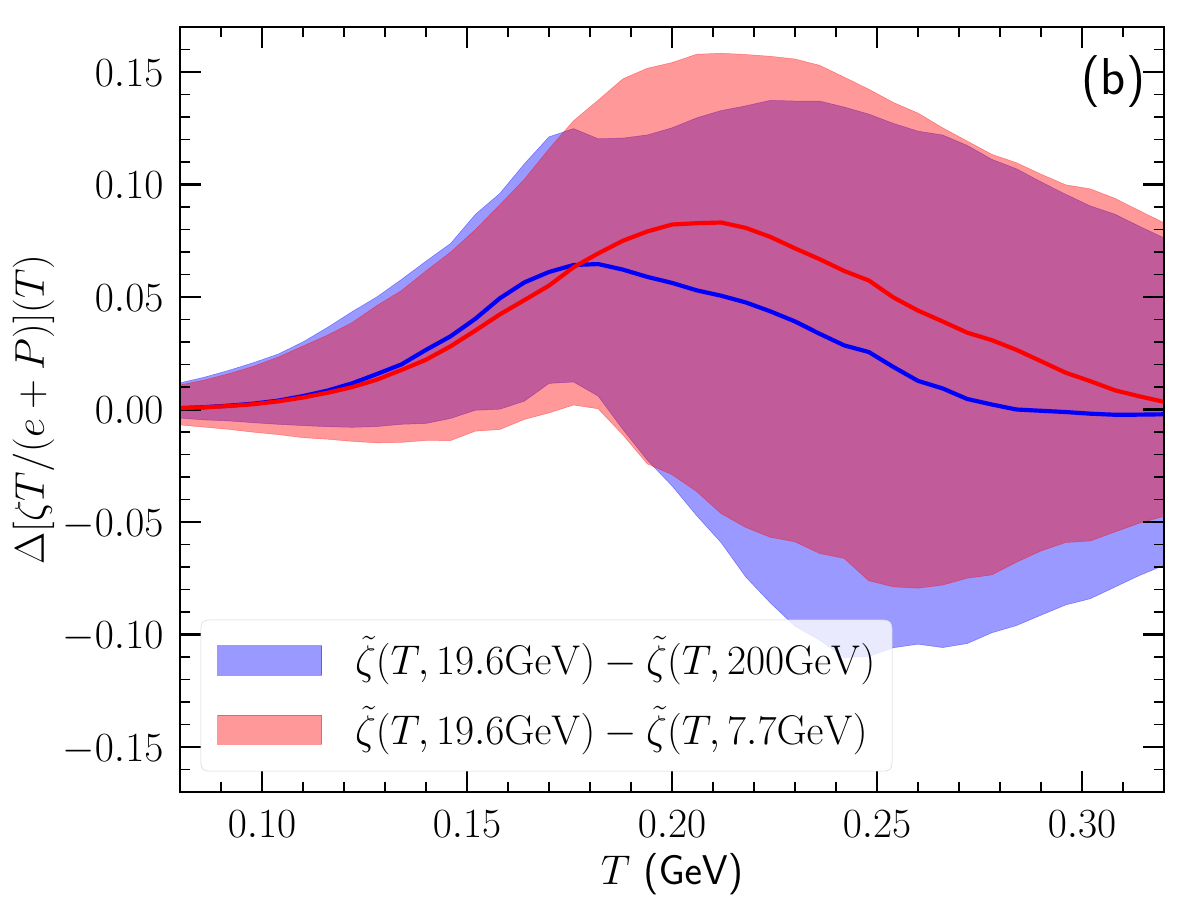}
    \caption{(Color online) Panel (a): Posterior distributions of the temperature dependence of the QGP $\tilde{\zeta}(T) = \frac{\zeta T}{(e+P)}(T)$ at different collision energies. Panel (b): Posterior distribution of the difference in $[\zeta T/(e+P)](T)$ at 19.6 GeV from the other two collision energies. Solid lines are the median of the $\Delta \tilde{\zeta}(T)$ distributions. Bands indicate 90\% confidence levels.}
    \label{fig:bulkPost}
\end{figure}
Figure~\ref{fig:shearPost} shows the posterior distribution for the effective QGP specific shear viscosity as a function of the net baryon chemical potential $\mu_B$. Using only the STAR mid-rapidity measurements at $\sqrt{s_\mathrm{NN}} = 200$ GeV in the Bayesian analysis constrains the effective QGP $\tilde{\eta} = \eta T/(e+P)$ around $\mu_B = 0$. The obtained 90\% posterior region is consistent with previous Bayesian analyses assuming longitudinal boost invariance~\cite{Bernhard:2016tnd,Bernhard:2019bmu,Nijs:2020ors,JETSCAPE:2020mzn,JETSCAPE:2020mzn,Parkkila:2021yha,Parkkila:2021tqq,Heffernan:2023gye,Heffernan:2023utr}. Notably, including PHOBOS pseudorapidity-dependent observables can significantly improve the constraints on the QGP shear viscosity up to $\mu_B$\,$\sim$\,0.2\,GeV. The sensitivity to $\eta T/(e + P)$ at finite $\mu_B$ comes from the fact that the fireballs in the forward and backward rapidity regions probe larger net baryon densities. Therefore, we emphasize that rapidity-dependent measurements at RHIC \cite{PHOBOS:2005zhy,PHOBOS:2006dbo} are extremely valuable to extract the $\mu_B$ dependence of the QGP properties. They have further been used to constrain the $T$ dependence of the shear viscosity in the low-temperature regime \cite{Denicol:2015nhu}, which in our framework is covered by the \URQMD\  simulations.

Finally, the Bayesian analysis including observables from the full RHIC BES program provides a significant constraint on the QGP $[\eta T/(e+P)](\mu_B)$ up to $\mu_B$\,$\sim$\,0.4\,GeV. 
We find that the RHIC BES measurements favor the QGP specific shear viscosity to \emph{increase} with $\mu_B$. This conclusion is consistent with previous phenomenological studies~\cite{Shen:2020jwv,Karpenko:2015xea,Auvinen:2017fjw} and calculations \cite{Hoyos:2020hmq,Soloveva:2020hpr,McLaughlin:2021dph}, but different from theoretical work in \cite{Denicol:2013nua, Grefa:2022sav}. Future studies including a more general ($T-\mu_B$) dependence of the shear viscosity will result in more robust constraints.

Figure~\ref{fig:bulkPost}a shows the posterior constraints on the QGP specific bulk viscosity $\tilde{\zeta}(T) \equiv [\zeta T/(e+P)] (T)$. The Bayesian analysis with only the measurements at 200 GeV favors a bulk viscosity peaking around $T = 200-220$\,MeV. The constraints at high temperature are relatively weak compared with the 90\% prior. The preferred values of $\tilde{\zeta}(T)$ at $\sqrt{s_\mathrm{NN}} =$ 19.6 GeV are larger than those at 200 and 7.7 GeV for temperatures between 0.15 GeV and 0.2 GeV. This non-monotonic behavior is further investigated in Fig.~\ref{fig:bulkPost}b, where we compute the difference $\Delta \tilde{\zeta}(T)$ between 19.6 GeV and the two other collision energies sample-by-sample drawn from the posterior. This treatment ensures that the 90\% confidence bands include the correlated variations of $\tilde{\zeta}(T)$ in different posterior samples. 

We find a bias of $\Delta \tilde{\zeta}(T)$ towards positive values for temperatures $T \in [0.15, 0.2]$\,GeV. Although the 90\% confidence bands cover $\Delta \tilde{\zeta}(T) = 0$, our result suggests a non-monotonic dependence of the QGP bulk viscosity along the net baryon chemical potential direction. Physically, this result could also emerge if there is a softening (and/or hardening) of the equation of state relative to the lattice-QCD-based \textsc{neos-bqs} around $\mu_B$\,$\sim$\,$0.2(0.4)$\,GeV. Our result is consistent with the theory expectation from the STAR two-pion interferometry analyses (often referred to as HBT radii) \cite{STAR:2014shf,Lacey:2014wqa}. Therefore, it is essential to include the HBT radii measurements in future Bayesian inference analyses~\cite{Pratt:2015zsa} to further improve the statistical significance of this result.

\paragraph{4. Conclusions.}

This work presented the first extraction of temperature and baryon chemical potential dependent QGP transport coefficients from a multi-system Bayesian inference study of particle production, mean transverse momentum, and flow anisotropy in the RHIC BES program using an event-by-event (3+1)D dynamical framework. Such a study requires large-scale computations, which only became possible recently with significant improvements in the numerical performance of the theoretical framework.

Using measurements from multiple collision energies, we obtained statistically robust constraints on initial-state nuclear stopping, $\mu_B$-dependent QGP shear viscosity, and the QGP bulk viscosity, including its effective $\mu_B$ dependence via its variation at different collision energies. Constraints on the average rapidity loss in the initial state are essential to quantitatively understand the longitudinal dynamics in these collisions, such as baryon and charge stopping and longitudinal flow decorrelation. The RHIC BES measurements favor a larger effective QGP specific shear viscosity at finite $\mu_B$ than at $\mu_B = 0$. This finding provides valuable insight when confronted with theoretical studies, which differ even qualitatively in the $\mu_B$ dependence of $\eta T/(e+P)$~\cite{Denicol:2013nua,Soloveva:2020hpr, Hoyos:2020hmq,Grefa:2022sav}.

We find a hint of non-monotonic dependence of the QGP specific bulk viscosity $\zeta T/(e+P)$ as a function of the collision energy. Because the bulk viscosity decelerates the local expansion, our finding could also indicate a softening of the equation of state for $\mu_B \sim 0.2$ GeV, and/or a hardening at $\mu_B \sim 0.4$ GeV, relative to the employed EOS. For a more conclusive result, a flexible equation of state with variable $\mu_B$ dependence should be included in the analysis. Further, the posterior constraint can be improved by introducing more experimental observables in the future.

Overall, our work marks a significant advancement in extracting QGP properties at finite net baryon density, using  systematic global analyses with  RHIC BES measurements. It paves the way to phenomenologically quantify the QCD phase diagram and search for a possible critical point and the associated first-order phase transition at large net baryon densities. It will be exciting to confront this theoretical framework with the upcoming RHIC BES phase II measurements and those from the future Facility for Antiproton and Ion Research (FAIR) in Europe.

\begin{acknowledgments}
We thank Robert Pisarski and Scott Pratt for useful discussions.
This work is supported by the U.S. Department of Energy, Office of Science, Office of Nuclear Physics, under DOE Contract No.~DE-SC0012704 (B.P.S.) and Award No.~DE-SC0021969 (C.S.).
C.S. acknowledges a DOE Office of Science Early Career Award. 
W.B.Z. is supported by the National Science Foundation (NSF) under grant numbers ACI-2004571 within the framework of the XSCAPE project of the JETSCAPE collaboration and US DOE under Contract No.~DE-AC02-05CH11231, and within the framework of the Saturated Glue (SURGE) Topical Theory Collaboration. This research was done using computational resources provided by the Open Science Grid (OSG)~\cite{Pordes:2007zzb, Sfiligoi:2009cct}, which is supported by the National Science Foundation award \#2030508.
\end{acknowledgments}

\bibliography{spires, non-spires}

\begin{thebibliography}{81}%
\makeatletter
\providecommand \@ifxundefined [1]{%
 \@ifx{#1\undefined}
}%
\providecommand \@ifnum [1]{%
 \ifnum #1\expandafter \@firstoftwo
 \else \expandafter \@secondoftwo
 \fi
}%
\providecommand \@ifx [1]{%
 \ifx #1\expandafter \@firstoftwo
 \else \expandafter \@secondoftwo
 \fi
}%
\providecommand \natexlab [1]{#1}%
\providecommand \enquote  [1]{``#1''}%
\providecommand \bibnamefont  [1]{#1}%
\providecommand \bibfnamefont [1]{#1}%
\providecommand \citenamefont [1]{#1}%
\providecommand \href@noop [0]{\@secondoftwo}%
\providecommand \href [0]{\begingroup \@sanitize@url \@href}%
\providecommand \@href[1]{\@@startlink{#1}\@@href}%
\providecommand \@@href[1]{\endgroup#1\@@endlink}%
\providecommand \@sanitize@url [0]{\catcode `\\12\catcode `\$12\catcode
  `\&12\catcode `\#12\catcode `\^12\catcode `\_12\catcode `\%12\relax}%
\providecommand \@@startlink[1]{}%
\providecommand \@@endlink[0]{}%
\providecommand \url  [0]{\begingroup\@sanitize@url \@url }%
\providecommand \@url [1]{\endgroup\@href {#1}{\urlprefix }}%
\providecommand \urlprefix  [0]{URL }%
\providecommand \Eprint [0]{\href }%
\providecommand \doibase [0]{http://dx.doi.org/}%
\providecommand \selectlanguage [0]{\@gobble}%
\providecommand \bibinfo  [0]{\@secondoftwo}%
\providecommand \bibfield  [0]{\@secondoftwo}%
\providecommand \translation [1]{[#1]}%
\providecommand \BibitemOpen [0]{}%
\providecommand \bibitemStop [0]{}%
\providecommand \bibitemNoStop [0]{.\EOS\space}%
\providecommand \EOS [0]{\spacefactor3000\relax}%
\providecommand \BibitemShut  [1]{\csname bibitem#1\endcsname}%
\let\auto@bib@innerbib\@empty
\bibitem [{\citenamefont {Achenbach}\ \emph {et~al.}(2023)\citenamefont
  {Achenbach} \emph {et~al.}}]{Achenbach:2023pba}%
  \BibitemOpen
  \bibfield  {author} {\bibinfo {author} {\bibfnamefont {P.}~\bibnamefont
  {Achenbach}} \emph {et~al.},\ }\bibfield  {title} {\enquote {\bibinfo {title}
  {{The Present and Future of QCD}},}\ }\href@noop {} {\  (\bibinfo {year}
  {2023})},\ \Eprint {http://arxiv.org/abs/2303.02579} {arXiv:2303.02579
  [hep-ph]} \BibitemShut {NoStop}%
\bibitem [{\citenamefont {Arslandok}\ \emph {et~al.}(2023)\citenamefont
  {Arslandok} \emph {et~al.}}]{Arslandok:2023utm}%
  \BibitemOpen
  \bibfield  {author} {\bibinfo {author} {\bibfnamefont {M.}~\bibnamefont
  {Arslandok}} \emph {et~al.},\ }\bibfield  {title} {\enquote {\bibinfo {title}
  {{Hot QCD White Paper}},}\ }\href@noop {} {\  (\bibinfo {year} {2023})},\
  \Eprint {http://arxiv.org/abs/2303.17254} {arXiv:2303.17254 [nucl-ex]}
  \BibitemShut {NoStop}%
\bibitem [{\citenamefont {Caines}(2009)}]{Caines:2009yu}%
  \BibitemOpen
  \bibfield  {author} {\bibinfo {author} {\bibfnamefont {Helen}\ \bibnamefont
  {Caines}} (\bibinfo {collaboration} {STAR}),\ }\bibfield  {title} {\enquote
  {\bibinfo {title} {{The RHIC Beam Energy Scan: STAR'S Perspective}},}\ }in\
  \href@noop {} {\emph {\bibinfo {booktitle} {{44th Rencontres de Moriond on
  QCD and High Energy Interactions}}}}\ (\bibinfo {year} {2009})\ pp.\ \bibinfo
  {pages} {375--378},\ \Eprint {http://arxiv.org/abs/0906.0305}
  {arXiv:0906.0305 [nucl-ex]} \BibitemShut {NoStop}%
\bibitem [{\citenamefont {Mohanty}(2011)}]{Mohanty:2011nm}%
  \BibitemOpen
  \bibfield  {author} {\bibinfo {author} {\bibfnamefont {Bedangadas}\
  \bibnamefont {Mohanty}} (\bibinfo {collaboration} {STAR}),\ }\bibfield
  {title} {\enquote {\bibinfo {title} {{STAR experiment results from the beam
  energy scan program at RHIC}},}\ }\href {\doibase
  10.1088/0954-3899/38/12/124023} {\bibfield  {journal} {\bibinfo  {journal}
  {J. Phys. G}\ }\textbf {\bibinfo {volume} {38}},\ \bibinfo {pages} {124023}
  (\bibinfo {year} {2011})},\ \Eprint {http://arxiv.org/abs/1106.5902}
  {arXiv:1106.5902 [nucl-ex]} \BibitemShut {NoStop}%
\bibitem [{\citenamefont {Mitchell}(2013)}]{Mitchell:2012mx}%
  \BibitemOpen
  \bibfield  {author} {\bibinfo {author} {\bibfnamefont {Jeffery~T.}\
  \bibnamefont {Mitchell}} (\bibinfo {collaboration} {PHENIX}),\ }\bibfield
  {title} {\enquote {\bibinfo {title} {{The RHIC Beam Energy Scan Program:
  Results from the PHENIX Experiment}},}\ }\href {\doibase
  10.1016/j.nuclphysa.2013.02.161} {\bibfield  {journal} {\bibinfo  {journal}
  {Nucl. Phys. A}\ }\textbf {\bibinfo {volume} {904-905}},\ \bibinfo {pages}
  {903c--906c} (\bibinfo {year} {2013})},\ \Eprint
  {http://arxiv.org/abs/1211.6139} {arXiv:1211.6139 [nucl-ex]} \BibitemShut
  {NoStop}%
\bibitem [{\citenamefont {Odyniec}(2015)}]{Odyniec:2015iaa}%
  \BibitemOpen
  \bibfield  {author} {\bibinfo {author} {\bibfnamefont {Grazyna}\ \bibnamefont
  {Odyniec}},\ }\bibfield  {title} {\enquote {\bibinfo {title} {{Future of the
  beam energy scan program at RHIC}},}\ }\href {\doibase
  10.1051/epjconf/20149503027} {\bibfield  {journal} {\bibinfo  {journal} {EPJ
  Web Conf.}\ }\textbf {\bibinfo {volume} {95}},\ \bibinfo {pages} {03027}
  (\bibinfo {year} {2015})}\BibitemShut {NoStop}%
\bibitem [{\citenamefont {Aggarwal}\ \emph {et~al.}(2010)\citenamefont
  {Aggarwal} \emph {et~al.}}]{STAR:2010vob}%
  \BibitemOpen
  \bibfield  {author} {\bibinfo {author} {\bibfnamefont {M.~M.}\ \bibnamefont
  {Aggarwal}} \emph {et~al.} (\bibinfo {collaboration} {STAR}),\ }\bibfield
  {title} {\enquote {\bibinfo {title} {{An Experimental Exploration of the QCD
  Phase Diagram: The Search for the Critical Point and the Onset of
  De-confinement}},}\ }\href@noop {} {\  (\bibinfo {year} {2010})},\ \Eprint
  {http://arxiv.org/abs/1007.2613} {arXiv:1007.2613 [nucl-ex]} \BibitemShut
  {NoStop}%
\bibitem [{\citenamefont {Luo}\ and\ \citenamefont {Xu}(2017)}]{Luo:2017faz}%
  \BibitemOpen
  \bibfield  {author} {\bibinfo {author} {\bibfnamefont {Xiaofeng}\
  \bibnamefont {Luo}}\ and\ \bibinfo {author} {\bibfnamefont {Nu}~\bibnamefont
  {Xu}},\ }\bibfield  {title} {\enquote {\bibinfo {title} {{Search for the QCD
  Critical Point with Fluctuations of Conserved Quantities in Relativistic
  Heavy-Ion Collisions at RHIC : An Overview}},}\ }\href {\doibase
  10.1007/s41365-017-0257-0} {\bibfield  {journal} {\bibinfo  {journal} {Nucl.
  Sci. Tech.}\ }\textbf {\bibinfo {volume} {28}},\ \bibinfo {pages} {112}
  (\bibinfo {year} {2017})},\ \Eprint {http://arxiv.org/abs/1701.02105}
  {arXiv:1701.02105 [nucl-ex]} \BibitemShut {NoStop}%
\bibitem [{\citenamefont {Bzdak}\ \emph {et~al.}(2020)\citenamefont {Bzdak},
  \citenamefont {Esumi}, \citenamefont {Koch}, \citenamefont {Liao},
  \citenamefont {Stephanov},\ and\ \citenamefont {Xu}}]{Bzdak:2019pkr}%
  \BibitemOpen
  \bibfield  {author} {\bibinfo {author} {\bibfnamefont {Adam}\ \bibnamefont
  {Bzdak}}, \bibinfo {author} {\bibfnamefont {Shinichi}\ \bibnamefont {Esumi}},
  \bibinfo {author} {\bibfnamefont {Volker}\ \bibnamefont {Koch}}, \bibinfo
  {author} {\bibfnamefont {Jinfeng}\ \bibnamefont {Liao}}, \bibinfo {author}
  {\bibfnamefont {Mikhail}\ \bibnamefont {Stephanov}}, \ and\ \bibinfo {author}
  {\bibfnamefont {Nu}~\bibnamefont {Xu}},\ }\bibfield  {title} {\enquote
  {\bibinfo {title} {{Mapping the Phases of Quantum Chromodynamics with Beam
  Energy Scan}},}\ }\href {\doibase 10.1016/j.physrep.2020.01.005} {\bibfield
  {journal} {\bibinfo  {journal} {Phys. Rept.}\ }\textbf {\bibinfo {volume}
  {853}},\ \bibinfo {pages} {1--87} (\bibinfo {year} {2020})},\ \Eprint
  {http://arxiv.org/abs/1906.00936} {arXiv:1906.00936 [nucl-th]} \BibitemShut
  {NoStop}%
\bibitem [{\citenamefont {An}\ \emph {et~al.}(2022)\citenamefont {An} \emph
  {et~al.}}]{An:2021wof}%
  \BibitemOpen
  \bibfield  {author} {\bibinfo {author} {\bibfnamefont {Xin}\ \bibnamefont
  {An}} \emph {et~al.},\ }\bibfield  {title} {\enquote {\bibinfo {title} {{The
  BEST framework for the search for the QCD critical point and the chiral
  magnetic effect}},}\ }\href {\doibase 10.1016/j.nuclphysa.2021.122343}
  {\bibfield  {journal} {\bibinfo  {journal} {Nucl. Phys. A}\ }\textbf
  {\bibinfo {volume} {1017}},\ \bibinfo {pages} {122343} (\bibinfo {year}
  {2022})},\ \Eprint {http://arxiv.org/abs/2108.13867} {arXiv:2108.13867
  [nucl-th]} \BibitemShut {NoStop}%
\bibitem [{\citenamefont {Shen}\ \emph {et~al.}(2016)\citenamefont {Shen},
  \citenamefont {Qiu}, \citenamefont {Song}, \citenamefont {Bernhard},
  \citenamefont {Bass},\ and\ \citenamefont {Heinz}}]{Shen:2014vra}%
  \BibitemOpen
  \bibfield  {author} {\bibinfo {author} {\bibfnamefont {Chun}\ \bibnamefont
  {Shen}}, \bibinfo {author} {\bibfnamefont {Zhi}\ \bibnamefont {Qiu}},
  \bibinfo {author} {\bibfnamefont {Huichao}\ \bibnamefont {Song}}, \bibinfo
  {author} {\bibfnamefont {Jonah}\ \bibnamefont {Bernhard}}, \bibinfo {author}
  {\bibfnamefont {Steffen}\ \bibnamefont {Bass}}, \ and\ \bibinfo {author}
  {\bibfnamefont {Ulrich}\ \bibnamefont {Heinz}},\ }\bibfield  {title}
  {\enquote {\bibinfo {title} {{The iEBE-VISHNU code package for relativistic
  heavy-ion collisions}},}\ }\href {\doibase 10.1016/j.cpc.2015.08.039}
  {\bibfield  {journal} {\bibinfo  {journal} {Comput. Phys. Commun.}\ }\textbf
  {\bibinfo {volume} {199}},\ \bibinfo {pages} {61--85} (\bibinfo {year}
  {2016})},\ \Eprint {http://arxiv.org/abs/1409.8164} {arXiv:1409.8164
  [nucl-th]} \BibitemShut {NoStop}%
\bibitem [{\citenamefont {Putschke}\ \emph {et~al.}(2019)\citenamefont
  {Putschke} \emph {et~al.}}]{Putschke:2019yrg}%
  \BibitemOpen
  \bibfield  {author} {\bibinfo {author} {\bibfnamefont {J.~H.}\ \bibnamefont
  {Putschke}} \emph {et~al.},\ }\bibfield  {title} {\enquote {\bibinfo {title}
  {{The JETSCAPE framework}},}\ }\href@noop {} {\  (\bibinfo {year} {2019})},\
  \Eprint {http://arxiv.org/abs/1903.07706} {arXiv:1903.07706 [nucl-th]}
  \BibitemShut {NoStop}%
\bibitem [{\citenamefont {Schenke}\ \emph
  {et~al.}(2020{\natexlab{a}})\citenamefont {Schenke}, \citenamefont {Shen},\
  and\ \citenamefont {Tribedy}}]{Schenke:2020mbo}%
  \BibitemOpen
  \bibfield  {author} {\bibinfo {author} {\bibfnamefont {Bjoern}\ \bibnamefont
  {Schenke}}, \bibinfo {author} {\bibfnamefont {Chun}\ \bibnamefont {Shen}}, \
  and\ \bibinfo {author} {\bibfnamefont {Prithwish}\ \bibnamefont {Tribedy}},\
  }\bibfield  {title} {\enquote {\bibinfo {title} {{Running the gamut of high
  energy nuclear collisions}},}\ }\href {\doibase 10.1103/PhysRevC.102.044905}
  {\bibfield  {journal} {\bibinfo  {journal} {Phys. Rev. C}\ }\textbf {\bibinfo
  {volume} {102}},\ \bibinfo {pages} {044905} (\bibinfo {year}
  {2020}{\natexlab{a}})},\ \Eprint {http://arxiv.org/abs/2005.14682}
  {arXiv:2005.14682 [nucl-th]} \BibitemShut {NoStop}%
\bibitem [{\citenamefont {Nijs}\ \emph
  {et~al.}(2021{\natexlab{a}})\citenamefont {Nijs}, \citenamefont {van~der
  Schee}, \citenamefont {G\"ursoy},\ and\ \citenamefont
  {Snellings}}]{Nijs:2020roc}%
  \BibitemOpen
  \bibfield  {author} {\bibinfo {author} {\bibfnamefont {Govert}\ \bibnamefont
  {Nijs}}, \bibinfo {author} {\bibfnamefont {Wilke}\ \bibnamefont {van~der
  Schee}}, \bibinfo {author} {\bibfnamefont {Umut}\ \bibnamefont {G\"ursoy}}, \
  and\ \bibinfo {author} {\bibfnamefont {Raimond}\ \bibnamefont {Snellings}},\
  }\bibfield  {title} {\enquote {\bibinfo {title} {{Bayesian analysis of heavy
  ion collisions with the heavy ion computational framework Trajectum}},}\
  }\href {\doibase 10.1103/PhysRevC.103.054909} {\bibfield  {journal} {\bibinfo
   {journal} {Phys. Rev. C}\ }\textbf {\bibinfo {volume} {103}},\ \bibinfo
  {pages} {054909} (\bibinfo {year} {2021}{\natexlab{a}})},\ \Eprint
  {http://arxiv.org/abs/2010.15134} {arXiv:2010.15134 [nucl-th]} \BibitemShut
  {NoStop}%
\bibitem [{\citenamefont {Pang}\ \emph {et~al.}(2018)\citenamefont {Pang},
  \citenamefont {Petersen},\ and\ \citenamefont {Wang}}]{Pang:2018zzo}%
  \BibitemOpen
  \bibfield  {author} {\bibinfo {author} {\bibfnamefont {Long-Gang}\
  \bibnamefont {Pang}}, \bibinfo {author} {\bibfnamefont {Hannah}\ \bibnamefont
  {Petersen}}, \ and\ \bibinfo {author} {\bibfnamefont {Xin-Nian}\ \bibnamefont
  {Wang}},\ }\bibfield  {title} {\enquote {\bibinfo {title} {{Pseudorapidity
  distribution and decorrelation of anisotropic flow within the
  open-computing-language implementation CLVisc hydrodynamics}},}\ }\href
  {\doibase 10.1103/PhysRevC.97.064918} {\bibfield  {journal} {\bibinfo
  {journal} {Phys. Rev. C}\ }\textbf {\bibinfo {volume} {97}},\ \bibinfo
  {pages} {064918} (\bibinfo {year} {2018})},\ \Eprint
  {http://arxiv.org/abs/1802.04449} {arXiv:1802.04449 [nucl-th]} \BibitemShut
  {NoStop}%
\bibitem [{\citenamefont {Song}\ \emph {et~al.}(2011)\citenamefont {Song},
  \citenamefont {Bass}, \citenamefont {Heinz}, \citenamefont {Hirano},\ and\
  \citenamefont {Shen}}]{Song:2010mg}%
  \BibitemOpen
  \bibfield  {author} {\bibinfo {author} {\bibfnamefont {Huichao}\ \bibnamefont
  {Song}}, \bibinfo {author} {\bibfnamefont {Steffen~A.}\ \bibnamefont {Bass}},
  \bibinfo {author} {\bibfnamefont {Ulrich}\ \bibnamefont {Heinz}}, \bibinfo
  {author} {\bibfnamefont {Tetsufumi}\ \bibnamefont {Hirano}}, \ and\ \bibinfo
  {author} {\bibfnamefont {Chun}\ \bibnamefont {Shen}},\ }\bibfield  {title}
  {\enquote {\bibinfo {title} {{200 A GeV Au+Au collisions serve a nearly
  perfect quark-gluon liquid}},}\ }\href {\doibase
  10.1103/PhysRevLett.106.192301} {\bibfield  {journal} {\bibinfo  {journal}
  {Phys. Rev. Lett.}\ }\textbf {\bibinfo {volume} {106}},\ \bibinfo {pages}
  {192301} (\bibinfo {year} {2011})},\ \bibinfo {note} {[Erratum:
  Phys.Rev.Lett. 109, 139904 (2012)]},\ \Eprint
  {http://arxiv.org/abs/1011.2783} {arXiv:1011.2783 [nucl-th]} \BibitemShut
  {NoStop}%
\bibitem [{\citenamefont {Schenke}\ \emph {et~al.}(2011)\citenamefont
  {Schenke}, \citenamefont {Jeon},\ and\ \citenamefont
  {Gale}}]{Schenke:2010rr}%
  \BibitemOpen
  \bibfield  {author} {\bibinfo {author} {\bibfnamefont {Bjorn}\ \bibnamefont
  {Schenke}}, \bibinfo {author} {\bibfnamefont {Sangyong}\ \bibnamefont
  {Jeon}}, \ and\ \bibinfo {author} {\bibfnamefont {Charles}\ \bibnamefont
  {Gale}},\ }\bibfield  {title} {\enquote {\bibinfo {title} {{Elliptic and
  triangular flow in event-by-event (3+1)D viscous hydrodynamics}},}\ }\href
  {\doibase 10.1103/PhysRevLett.106.042301} {\bibfield  {journal} {\bibinfo
  {journal} {Phys. Rev. Lett.}\ }\textbf {\bibinfo {volume} {106}},\ \bibinfo
  {pages} {042301} (\bibinfo {year} {2011})},\ \Eprint
  {http://arxiv.org/abs/1009.3244} {arXiv:1009.3244 [hep-ph]} \BibitemShut
  {NoStop}%
\bibitem [{\citenamefont {Qiu}\ \emph {et~al.}(2012)\citenamefont {Qiu},
  \citenamefont {Shen},\ and\ \citenamefont {Heinz}}]{Qiu:2011hf}%
  \BibitemOpen
  \bibfield  {author} {\bibinfo {author} {\bibfnamefont {Zhi}\ \bibnamefont
  {Qiu}}, \bibinfo {author} {\bibfnamefont {Chun}\ \bibnamefont {Shen}}, \ and\
  \bibinfo {author} {\bibfnamefont {Ulrich}\ \bibnamefont {Heinz}},\ }\bibfield
   {title} {\enquote {\bibinfo {title} {{Hydrodynamic elliptic and triangular
  flow in Pb-Pb collisions at $\sqrt{s}=2.76$ATeV}},}\ }\href {\doibase
  10.1016/j.physletb.2011.12.041} {\bibfield  {journal} {\bibinfo  {journal}
  {Phys. Lett. B}\ }\textbf {\bibinfo {volume} {707}},\ \bibinfo {pages}
  {151--155} (\bibinfo {year} {2012})},\ \Eprint
  {http://arxiv.org/abs/1110.3033} {arXiv:1110.3033 [nucl-th]} \BibitemShut
  {NoStop}%
\bibitem [{\citenamefont {Gale}\ \emph {et~al.}(2013)\citenamefont {Gale},
  \citenamefont {Jeon},\ and\ \citenamefont {Schenke}}]{Gale:2013da}%
  \BibitemOpen
  \bibfield  {author} {\bibinfo {author} {\bibfnamefont {Charles}\ \bibnamefont
  {Gale}}, \bibinfo {author} {\bibfnamefont {Sangyong}\ \bibnamefont {Jeon}}, \
  and\ \bibinfo {author} {\bibfnamefont {Bjoern}\ \bibnamefont {Schenke}},\
  }\bibfield  {title} {\enquote {\bibinfo {title} {{Hydrodynamic Modeling of
  Heavy-Ion Collisions}},}\ }\href {\doibase 10.1142/S0217751X13400113}
  {\bibfield  {journal} {\bibinfo  {journal} {Int. J. Mod. Phys. A}\ }\textbf
  {\bibinfo {volume} {28}},\ \bibinfo {pages} {1340011} (\bibinfo {year}
  {2013})},\ \Eprint {http://arxiv.org/abs/1301.5893} {arXiv:1301.5893
  [nucl-th]} \BibitemShut {NoStop}%
\bibitem [{\citenamefont {Heinz}\ and\ \citenamefont
  {Snellings}(2013)}]{Heinz:2013th}%
  \BibitemOpen
  \bibfield  {author} {\bibinfo {author} {\bibfnamefont {Ulrich}\ \bibnamefont
  {Heinz}}\ and\ \bibinfo {author} {\bibfnamefont {Raimond}\ \bibnamefont
  {Snellings}},\ }\bibfield  {title} {\enquote {\bibinfo {title} {{Collective
  flow and viscosity in relativistic heavy-ion collisions}},}\ }\href {\doibase
  10.1146/annurev-nucl-102212-170540} {\bibfield  {journal} {\bibinfo
  {journal} {Ann. Rev. Nucl. Part. Sci.}\ }\textbf {\bibinfo {volume} {63}},\
  \bibinfo {pages} {123--151} (\bibinfo {year} {2013})},\ \Eprint
  {http://arxiv.org/abs/1301.2826} {arXiv:1301.2826 [nucl-th]} \BibitemShut
  {NoStop}%
\bibitem [{\citenamefont {Schenke}\ \emph
  {et~al.}(2020{\natexlab{b}})\citenamefont {Schenke}, \citenamefont {Shen},\
  and\ \citenamefont {Tribedy}}]{Schenke:2019pmk}%
  \BibitemOpen
  \bibfield  {author} {\bibinfo {author} {\bibfnamefont {Bjoern}\ \bibnamefont
  {Schenke}}, \bibinfo {author} {\bibfnamefont {Chun}\ \bibnamefont {Shen}}, \
  and\ \bibinfo {author} {\bibfnamefont {Prithwish}\ \bibnamefont {Tribedy}},\
  }\bibfield  {title} {\enquote {\bibinfo {title} {{Hybrid Color Glass
  Condensate and hydrodynamic description of the Relativistic Heavy Ion
  Collider small system scan}},}\ }\href {\doibase
  10.1016/j.physletb.2020.135322} {\bibfield  {journal} {\bibinfo  {journal}
  {Phys. Lett. B}\ }\textbf {\bibinfo {volume} {803}},\ \bibinfo {pages}
  {135322} (\bibinfo {year} {2020}{\natexlab{b}})},\ \Eprint
  {http://arxiv.org/abs/1908.06212} {arXiv:1908.06212 [nucl-th]} \BibitemShut
  {NoStop}%
\bibitem [{\citenamefont {Shen}\ and\ \citenamefont
  {Yan}(2020)}]{Shen:2020mgh}%
  \BibitemOpen
  \bibfield  {author} {\bibinfo {author} {\bibfnamefont {Chun}\ \bibnamefont
  {Shen}}\ and\ \bibinfo {author} {\bibfnamefont {Li}~\bibnamefont {Yan}},\
  }\bibfield  {title} {\enquote {\bibinfo {title} {{Recent development of
  hydrodynamic modeling in heavy-ion collisions}},}\ }\href {\doibase
  10.1007/s41365-020-00829-z} {\bibfield  {journal} {\bibinfo  {journal} {Nucl.
  Sci. Tech.}\ }\textbf {\bibinfo {volume} {31}},\ \bibinfo {pages} {122}
  (\bibinfo {year} {2020})},\ \Eprint {http://arxiv.org/abs/2010.12377}
  {arXiv:2010.12377 [nucl-th]} \BibitemShut {NoStop}%
\bibitem [{\citenamefont {Altenkort}\ \emph {et~al.}(2023)\citenamefont
  {Altenkort}, \citenamefont {Eller}, \citenamefont {Francis}, \citenamefont
  {Kaczmarek}, \citenamefont {Mazur}, \citenamefont {Moore},\ and\
  \citenamefont {Shu}}]{Altenkort:2022yhb}%
  \BibitemOpen
  \bibfield  {author} {\bibinfo {author} {\bibfnamefont {Luis}\ \bibnamefont
  {Altenkort}}, \bibinfo {author} {\bibfnamefont {Alexander~M.}\ \bibnamefont
  {Eller}}, \bibinfo {author} {\bibfnamefont {Anthony}\ \bibnamefont
  {Francis}}, \bibinfo {author} {\bibfnamefont {Olaf}\ \bibnamefont
  {Kaczmarek}}, \bibinfo {author} {\bibfnamefont {Lukas}\ \bibnamefont
  {Mazur}}, \bibinfo {author} {\bibfnamefont {Guy~D.}\ \bibnamefont {Moore}}, \
  and\ \bibinfo {author} {\bibfnamefont {Hai-Tao}\ \bibnamefont {Shu}},\
  }\bibfield  {title} {\enquote {\bibinfo {title} {{Viscosity of pure-glue QCD
  from the lattice}},}\ }\href {\doibase 10.1103/PhysRevD.108.014503}
  {\bibfield  {journal} {\bibinfo  {journal} {Phys. Rev. D}\ }\textbf {\bibinfo
  {volume} {108}},\ \bibinfo {pages} {014503} (\bibinfo {year} {2023})},\
  \Eprint {http://arxiv.org/abs/2211.08230} {arXiv:2211.08230 [hep-lat]}
  \BibitemShut {NoStop}%
\bibitem [{\citenamefont {Ryu}\ \emph {et~al.}(2015)\citenamefont {Ryu},
  \citenamefont {Paquet}, \citenamefont {Shen}, \citenamefont {Denicol},
  \citenamefont {Schenke}, \citenamefont {Jeon},\ and\ \citenamefont
  {Gale}}]{Ryu:2015vwa}%
  \BibitemOpen
  \bibfield  {author} {\bibinfo {author} {\bibfnamefont {S.}~\bibnamefont
  {Ryu}}, \bibinfo {author} {\bibfnamefont {J.~F.}\ \bibnamefont {Paquet}},
  \bibinfo {author} {\bibfnamefont {C.}~\bibnamefont {Shen}}, \bibinfo {author}
  {\bibfnamefont {G.~S.}\ \bibnamefont {Denicol}}, \bibinfo {author}
  {\bibfnamefont {B.}~\bibnamefont {Schenke}}, \bibinfo {author} {\bibfnamefont
  {S.}~\bibnamefont {Jeon}}, \ and\ \bibinfo {author} {\bibfnamefont
  {C.}~\bibnamefont {Gale}},\ }\bibfield  {title} {\enquote {\bibinfo {title}
  {{Importance of the Bulk Viscosity of QCD in Ultrarelativistic Heavy-Ion
  Collisions}},}\ }\href {\doibase 10.1103/PhysRevLett.115.132301} {\bibfield
  {journal} {\bibinfo  {journal} {Phys. Rev. Lett.}\ }\textbf {\bibinfo
  {volume} {115}},\ \bibinfo {pages} {132301} (\bibinfo {year} {2015})},\
  \Eprint {http://arxiv.org/abs/1502.01675} {arXiv:1502.01675 [nucl-th]}
  \BibitemShut {NoStop}%
\bibitem [{\citenamefont {Karpenko}\ \emph {et~al.}(2015)\citenamefont
  {Karpenko}, \citenamefont {Huovinen}, \citenamefont {Petersen},\ and\
  \citenamefont {Bleicher}}]{Karpenko:2015xea}%
  \BibitemOpen
  \bibfield  {author} {\bibinfo {author} {\bibfnamefont {Iu.~A.}\ \bibnamefont
  {Karpenko}}, \bibinfo {author} {\bibfnamefont {P.}~\bibnamefont {Huovinen}},
  \bibinfo {author} {\bibfnamefont {H.}~\bibnamefont {Petersen}}, \ and\
  \bibinfo {author} {\bibfnamefont {M.}~\bibnamefont {Bleicher}},\ }\bibfield
  {title} {\enquote {\bibinfo {title} {{Estimation of the shear viscosity at
  finite net-baryon density from $A+A$ collision data at $\sqrt{s_\mathrm{NN}}
  = 7.7-200$ GeV}},}\ }\href {\doibase 10.1103/PhysRevC.91.064901} {\bibfield
  {journal} {\bibinfo  {journal} {Phys. Rev. C}\ }\textbf {\bibinfo {volume}
  {91}},\ \bibinfo {pages} {064901} (\bibinfo {year} {2015})},\ \Eprint
  {http://arxiv.org/abs/1502.01978} {arXiv:1502.01978 [nucl-th]} \BibitemShut
  {NoStop}%
\bibitem [{\citenamefont {Shen}\ and\ \citenamefont
  {Heinz}(2015)}]{Shen:2015msa}%
  \BibitemOpen
  \bibfield  {author} {\bibinfo {author} {\bibfnamefont {Chun}\ \bibnamefont
  {Shen}}\ and\ \bibinfo {author} {\bibfnamefont {Ulrich}\ \bibnamefont
  {Heinz}},\ }\bibfield  {title} {\enquote {\bibinfo {title} {{The road to
  precision: Extraction of the specific shear viscosity of the quark-gluon
  plasma}},}\ }\href {\doibase 10.1080/10619127.2015.1006502} {\bibfield
  {journal} {\bibinfo  {journal} {Nucl. Phys. News}\ }\textbf {\bibinfo
  {volume} {25}},\ \bibinfo {pages} {6--11} (\bibinfo {year} {2015})},\ \Eprint
  {http://arxiv.org/abs/1507.01558} {arXiv:1507.01558 [nucl-th]} \BibitemShut
  {NoStop}%
\bibitem [{\citenamefont {Ryu}\ \emph {et~al.}(2018)\citenamefont {Ryu},
  \citenamefont {Paquet}, \citenamefont {Shen}, \citenamefont {Denicol},
  \citenamefont {Schenke}, \citenamefont {Jeon},\ and\ \citenamefont
  {Gale}}]{Ryu:2017qzn}%
  \BibitemOpen
  \bibfield  {author} {\bibinfo {author} {\bibfnamefont {Sangwook}\
  \bibnamefont {Ryu}}, \bibinfo {author} {\bibfnamefont {Jean-Fran\c{c}ois}\
  \bibnamefont {Paquet}}, \bibinfo {author} {\bibfnamefont {Chun}\ \bibnamefont
  {Shen}}, \bibinfo {author} {\bibfnamefont {Gabriel}\ \bibnamefont {Denicol}},
  \bibinfo {author} {\bibfnamefont {Bj\"orn}\ \bibnamefont {Schenke}}, \bibinfo
  {author} {\bibfnamefont {Sangyong}\ \bibnamefont {Jeon}}, \ and\ \bibinfo
  {author} {\bibfnamefont {Charles}\ \bibnamefont {Gale}},\ }\bibfield  {title}
  {\enquote {\bibinfo {title} {{Effects of bulk viscosity and hadronic
  rescattering in heavy ion collisions at energies available at the BNL
  Relativistic Heavy Ion Collider and at the CERN Large Hadron Collider}},}\
  }\href {\doibase 10.1103/PhysRevC.97.034910} {\bibfield  {journal} {\bibinfo
  {journal} {Phys. Rev. C}\ }\textbf {\bibinfo {volume} {97}},\ \bibinfo
  {pages} {034910} (\bibinfo {year} {2018})},\ \Eprint
  {http://arxiv.org/abs/1704.04216} {arXiv:1704.04216 [nucl-th]} \BibitemShut
  {NoStop}%
\bibitem [{\citenamefont {Schenke}\ \emph {et~al.}(2019)\citenamefont
  {Schenke}, \citenamefont {Shen},\ and\ \citenamefont
  {Tribedy}}]{Schenke:2019ruo}%
  \BibitemOpen
  \bibfield  {author} {\bibinfo {author} {\bibfnamefont {Bj\"orn}\ \bibnamefont
  {Schenke}}, \bibinfo {author} {\bibfnamefont {Chun}\ \bibnamefont {Shen}}, \
  and\ \bibinfo {author} {\bibfnamefont {Prithwish}\ \bibnamefont {Tribedy}},\
  }\bibfield  {title} {\enquote {\bibinfo {title} {{Multi-particle and
  charge-dependent azimuthal correlations in heavy-ion collisions at the
  Relativistic Heavy-Ion Collider}},}\ }\href {\doibase
  10.1103/PhysRevC.99.044908} {\bibfield  {journal} {\bibinfo  {journal} {Phys.
  Rev. C}\ }\textbf {\bibinfo {volume} {99}},\ \bibinfo {pages} {044908}
  (\bibinfo {year} {2019})},\ \Eprint {http://arxiv.org/abs/1901.04378}
  {arXiv:1901.04378 [nucl-th]} \BibitemShut {NoStop}%
\bibitem [{\citenamefont {Niemi}\ \emph {et~al.}(2016)\citenamefont {Niemi},
  \citenamefont {Eskola},\ and\ \citenamefont {Paatelainen}}]{Niemi:2015qia}%
  \BibitemOpen
  \bibfield  {author} {\bibinfo {author} {\bibfnamefont {H.}~\bibnamefont
  {Niemi}}, \bibinfo {author} {\bibfnamefont {K.~J.}\ \bibnamefont {Eskola}}, \
  and\ \bibinfo {author} {\bibfnamefont {R.}~\bibnamefont {Paatelainen}},\
  }\bibfield  {title} {\enquote {\bibinfo {title} {{Event-by-event fluctuations
  in a perturbative QCD + saturation + hydrodynamics model: Determining QCD
  matter shear viscosity in ultrarelativistic heavy-ion collisions}},}\ }\href
  {\doibase 10.1103/PhysRevC.93.024907} {\bibfield  {journal} {\bibinfo
  {journal} {Phys. Rev. C}\ }\textbf {\bibinfo {volume} {93}},\ \bibinfo
  {pages} {024907} (\bibinfo {year} {2016})},\ \Eprint
  {http://arxiv.org/abs/1505.02677} {arXiv:1505.02677 [hep-ph]} \BibitemShut
  {NoStop}%
\bibitem [{\citenamefont {Shen}\ and\ \citenamefont
  {Alzhrani}(2020)}]{Shen:2020jwv}%
  \BibitemOpen
  \bibfield  {author} {\bibinfo {author} {\bibfnamefont {Chun}\ \bibnamefont
  {Shen}}\ and\ \bibinfo {author} {\bibfnamefont {Sahr}\ \bibnamefont
  {Alzhrani}},\ }\bibfield  {title} {\enquote {\bibinfo {title}
  {{Collision-geometry-based 3D initial condition for relativistic heavy-ion
  collisions}},}\ }\href {\doibase 10.1103/PhysRevC.102.014909} {\bibfield
  {journal} {\bibinfo  {journal} {Phys. Rev. C}\ }\textbf {\bibinfo {volume}
  {102}},\ \bibinfo {pages} {014909} (\bibinfo {year} {2020})},\ \Eprint
  {http://arxiv.org/abs/2003.05852} {arXiv:2003.05852 [nucl-th]} \BibitemShut
  {NoStop}%
\bibitem [{\citenamefont {Pratt}\ \emph {et~al.}(2015)\citenamefont {Pratt},
  \citenamefont {Sangaline}, \citenamefont {Sorensen},\ and\ \citenamefont
  {Wang}}]{Pratt:2015zsa}%
  \BibitemOpen
  \bibfield  {author} {\bibinfo {author} {\bibfnamefont {Scott}\ \bibnamefont
  {Pratt}}, \bibinfo {author} {\bibfnamefont {Evan}\ \bibnamefont {Sangaline}},
  \bibinfo {author} {\bibfnamefont {Paul}\ \bibnamefont {Sorensen}}, \ and\
  \bibinfo {author} {\bibfnamefont {Hui}\ \bibnamefont {Wang}},\ }\bibfield
  {title} {\enquote {\bibinfo {title} {{Constraining the Eq. of State of
  Super-Hadronic Matter from Heavy-Ion Collisions}},}\ }\href {\doibase
  10.1103/PhysRevLett.114.202301} {\bibfield  {journal} {\bibinfo  {journal}
  {Phys. Rev. Lett.}\ }\textbf {\bibinfo {volume} {114}},\ \bibinfo {pages}
  {202301} (\bibinfo {year} {2015})},\ \Eprint
  {http://arxiv.org/abs/1501.04042} {arXiv:1501.04042 [nucl-th]} \BibitemShut
  {NoStop}%
\bibitem [{\citenamefont {Bernhard}\ \emph {et~al.}(2016)\citenamefont
  {Bernhard}, \citenamefont {Moreland}, \citenamefont {Bass}, \citenamefont
  {Liu},\ and\ \citenamefont {Heinz}}]{Bernhard:2016tnd}%
  \BibitemOpen
  \bibfield  {author} {\bibinfo {author} {\bibfnamefont {Jonah~E.}\
  \bibnamefont {Bernhard}}, \bibinfo {author} {\bibfnamefont {J.~Scott}\
  \bibnamefont {Moreland}}, \bibinfo {author} {\bibfnamefont {Steffen~A.}\
  \bibnamefont {Bass}}, \bibinfo {author} {\bibfnamefont {Jia}\ \bibnamefont
  {Liu}}, \ and\ \bibinfo {author} {\bibfnamefont {Ulrich}\ \bibnamefont
  {Heinz}},\ }\bibfield  {title} {\enquote {\bibinfo {title} {{Applying
  Bayesian parameter estimation to relativistic heavy-ion collisions:
  simultaneous characterization of the initial state and quark-gluon plasma
  medium}},}\ }\href {\doibase 10.1103/PhysRevC.94.024907} {\bibfield
  {journal} {\bibinfo  {journal} {Phys. Rev. C}\ }\textbf {\bibinfo {volume}
  {94}},\ \bibinfo {pages} {024907} (\bibinfo {year} {2016})},\ \Eprint
  {http://arxiv.org/abs/1605.03954} {arXiv:1605.03954 [nucl-th]} \BibitemShut
  {NoStop}%
\bibitem [{\citenamefont {Auvinen}\ \emph {et~al.}(2018)\citenamefont
  {Auvinen}, \citenamefont {Bernhard}, \citenamefont {Bass},\ and\
  \citenamefont {Karpenko}}]{Auvinen:2017fjw}%
  \BibitemOpen
  \bibfield  {author} {\bibinfo {author} {\bibfnamefont {Jussi}\ \bibnamefont
  {Auvinen}}, \bibinfo {author} {\bibfnamefont {Jonah~E.}\ \bibnamefont
  {Bernhard}}, \bibinfo {author} {\bibfnamefont {Steffen~A.}\ \bibnamefont
  {Bass}}, \ and\ \bibinfo {author} {\bibfnamefont {Iurii}\ \bibnamefont
  {Karpenko}},\ }\bibfield  {title} {\enquote {\bibinfo {title} {{Investigating
  the collision energy dependence of $\eta$/s in the beam energy scan at the
  BNL Relativistic Heavy Ion Collider using Bayesian statistics}},}\ }\href
  {\doibase 10.1103/PhysRevC.97.044905} {\bibfield  {journal} {\bibinfo
  {journal} {Phys. Rev. C}\ }\textbf {\bibinfo {volume} {97}},\ \bibinfo
  {pages} {044905} (\bibinfo {year} {2018})},\ \Eprint
  {http://arxiv.org/abs/1706.03666} {arXiv:1706.03666 [hep-ph]} \BibitemShut
  {NoStop}%
\bibitem [{\citenamefont {Bernhard}\ \emph {et~al.}(2019)\citenamefont
  {Bernhard}, \citenamefont {Moreland},\ and\ \citenamefont
  {Bass}}]{Bernhard:2019bmu}%
  \BibitemOpen
  \bibfield  {author} {\bibinfo {author} {\bibfnamefont {Jonah~E.}\
  \bibnamefont {Bernhard}}, \bibinfo {author} {\bibfnamefont {J.~Scott}\
  \bibnamefont {Moreland}}, \ and\ \bibinfo {author} {\bibfnamefont
  {Steffen~A.}\ \bibnamefont {Bass}},\ }\bibfield  {title} {\enquote {\bibinfo
  {title} {{Bayesian estimation of the specific shear and bulk viscosity of
  quark\textendash{}gluon plasma}},}\ }\href {\doibase
  10.1038/s41567-019-0611-8} {\bibfield  {journal} {\bibinfo  {journal} {Nature
  Phys.}\ }\textbf {\bibinfo {volume} {15}},\ \bibinfo {pages} {1113--1117}
  (\bibinfo {year} {2019})}\BibitemShut {NoStop}%
\bibitem [{\citenamefont {Nijs}\ \emph
  {et~al.}(2021{\natexlab{b}})\citenamefont {Nijs}, \citenamefont {van~der
  Schee}, \citenamefont {G\"ursoy},\ and\ \citenamefont
  {Snellings}}]{Nijs:2020ors}%
  \BibitemOpen
  \bibfield  {author} {\bibinfo {author} {\bibfnamefont {Govert}\ \bibnamefont
  {Nijs}}, \bibinfo {author} {\bibfnamefont {Wilke}\ \bibnamefont {van~der
  Schee}}, \bibinfo {author} {\bibfnamefont {Umut}\ \bibnamefont {G\"ursoy}}, \
  and\ \bibinfo {author} {\bibfnamefont {Raimond}\ \bibnamefont {Snellings}},\
  }\bibfield  {title} {\enquote {\bibinfo {title} {{Transverse Momentum
  Differential Global Analysis of Heavy-Ion Collisions}},}\ }\href {\doibase
  10.1103/PhysRevLett.126.202301} {\bibfield  {journal} {\bibinfo  {journal}
  {Phys. Rev. Lett.}\ }\textbf {\bibinfo {volume} {126}},\ \bibinfo {pages}
  {202301} (\bibinfo {year} {2021}{\natexlab{b}})},\ \Eprint
  {http://arxiv.org/abs/2010.15130} {arXiv:2010.15130 [nucl-th]} \BibitemShut
  {NoStop}%
\bibitem [{\citenamefont {Everett}\ \emph
  {et~al.}(2021{\natexlab{a}})\citenamefont {Everett} \emph
  {et~al.}}]{JETSCAPE:2020shq}%
  \BibitemOpen
  \bibfield  {author} {\bibinfo {author} {\bibfnamefont {D.}~\bibnamefont
  {Everett}} \emph {et~al.} (\bibinfo {collaboration} {JETSCAPE}),\ }\bibfield
  {title} {\enquote {\bibinfo {title} {{Phenomenological constraints on the
  transport properties of QCD matter with data-driven model averaging}},}\
  }\href {\doibase 10.1103/PhysRevLett.126.242301} {\bibfield  {journal}
  {\bibinfo  {journal} {Phys. Rev. Lett.}\ }\textbf {\bibinfo {volume} {126}},\
  \bibinfo {pages} {242301} (\bibinfo {year} {2021}{\natexlab{a}})},\ \Eprint
  {http://arxiv.org/abs/2010.03928} {arXiv:2010.03928 [hep-ph]} \BibitemShut
  {NoStop}%
\bibitem [{\citenamefont {Everett}\ \emph
  {et~al.}(2021{\natexlab{b}})\citenamefont {Everett} \emph
  {et~al.}}]{JETSCAPE:2020mzn}%
  \BibitemOpen
  \bibfield  {author} {\bibinfo {author} {\bibfnamefont {D.}~\bibnamefont
  {Everett}} \emph {et~al.} (\bibinfo {collaboration} {JETSCAPE}),\ }\bibfield
  {title} {\enquote {\bibinfo {title} {{Multisystem Bayesian constraints on the
  transport coefficients of QCD matter}},}\ }\href {\doibase
  10.1103/PhysRevC.103.054904} {\bibfield  {journal} {\bibinfo  {journal}
  {Phys. Rev. C}\ }\textbf {\bibinfo {volume} {103}},\ \bibinfo {pages}
  {054904} (\bibinfo {year} {2021}{\natexlab{b}})},\ \Eprint
  {http://arxiv.org/abs/2011.01430} {arXiv:2011.01430 [hep-ph]} \BibitemShut
  {NoStop}%
\bibitem [{\citenamefont {Parkkila}\ \emph {et~al.}(2022)\citenamefont
  {Parkkila}, \citenamefont {Onnerstad}, \citenamefont {Taghavi}, \citenamefont
  {Mordasini}, \citenamefont {Bilandzic}, \citenamefont {Virta},\ and\
  \citenamefont {Kim}}]{Parkkila:2021yha}%
  \BibitemOpen
  \bibfield  {author} {\bibinfo {author} {\bibfnamefont {J.~E.}\ \bibnamefont
  {Parkkila}}, \bibinfo {author} {\bibfnamefont {A.}~\bibnamefont {Onnerstad}},
  \bibinfo {author} {\bibfnamefont {S.~F.}\ \bibnamefont {Taghavi}}, \bibinfo
  {author} {\bibfnamefont {C.}~\bibnamefont {Mordasini}}, \bibinfo {author}
  {\bibfnamefont {A.}~\bibnamefont {Bilandzic}}, \bibinfo {author}
  {\bibfnamefont {M.}~\bibnamefont {Virta}}, \ and\ \bibinfo {author}
  {\bibfnamefont {D.~J.}\ \bibnamefont {Kim}},\ }\bibfield  {title} {\enquote
  {\bibinfo {title} {{New constraints for QCD matter from improved Bayesian
  parameter estimation in heavy-ion collisions at LHC}},}\ }\href {\doibase
  10.1016/j.physletb.2022.137485} {\bibfield  {journal} {\bibinfo  {journal}
  {Phys. Lett. B}\ }\textbf {\bibinfo {volume} {835}},\ \bibinfo {pages}
  {137485} (\bibinfo {year} {2022})},\ \Eprint
  {http://arxiv.org/abs/2111.08145} {arXiv:2111.08145 [hep-ph]} \BibitemShut
  {NoStop}%
\bibitem [{\citenamefont {Parkkila}\ \emph {et~al.}(2021)\citenamefont
  {Parkkila}, \citenamefont {Onnerstad},\ and\ \citenamefont
  {Kim}}]{Parkkila:2021tqq}%
  \BibitemOpen
  \bibfield  {author} {\bibinfo {author} {\bibfnamefont {J.~E.}\ \bibnamefont
  {Parkkila}}, \bibinfo {author} {\bibfnamefont {A.}~\bibnamefont {Onnerstad}},
  \ and\ \bibinfo {author} {\bibfnamefont {D.~J.}\ \bibnamefont {Kim}},\
  }\bibfield  {title} {\enquote {\bibinfo {title} {{Bayesian estimation of the
  specific shear and bulk viscosity of the quark-gluon plasma with additional
  flow harmonic observables}},}\ }\href {\doibase 10.1103/PhysRevC.104.054904}
  {\bibfield  {journal} {\bibinfo  {journal} {Phys. Rev. C}\ }\textbf {\bibinfo
  {volume} {104}},\ \bibinfo {pages} {054904} (\bibinfo {year} {2021})},\
  \Eprint {http://arxiv.org/abs/2106.05019} {arXiv:2106.05019 [hep-ph]}
  \BibitemShut {NoStop}%
\bibitem [{\citenamefont {Phillips}\ \emph {et~al.}(2021)\citenamefont
  {Phillips} \emph {et~al.}}]{Phillips:2020dmw}%
  \BibitemOpen
  \bibfield  {author} {\bibinfo {author} {\bibfnamefont {D.~R.}\ \bibnamefont
  {Phillips}} \emph {et~al.},\ }\bibfield  {title} {\enquote {\bibinfo {title}
  {{Get on the BAND Wagon: A Bayesian Framework for Quantifying Model
  Uncertainties in Nuclear Dynamics}},}\ }\href {\doibase
  10.1088/1361-6471/abf1df} {\bibfield  {journal} {\bibinfo  {journal} {J.
  Phys. G}\ }\textbf {\bibinfo {volume} {48}},\ \bibinfo {pages} {072001}
  (\bibinfo {year} {2021})},\ \Eprint {http://arxiv.org/abs/2012.07704}
  {arXiv:2012.07704 [nucl-th]} \BibitemShut {NoStop}%
\bibitem [{\citenamefont {Heffernan}\ \emph
  {et~al.}(2023{\natexlab{a}})\citenamefont {Heffernan}, \citenamefont {Gale},
  \citenamefont {Jeon},\ and\ \citenamefont {Paquet}}]{Heffernan:2023gye}%
  \BibitemOpen
  \bibfield  {author} {\bibinfo {author} {\bibfnamefont {Matthew~R.}\
  \bibnamefont {Heffernan}}, \bibinfo {author} {\bibfnamefont {Charles}\
  \bibnamefont {Gale}}, \bibinfo {author} {\bibfnamefont {Sangyong}\
  \bibnamefont {Jeon}}, \ and\ \bibinfo {author} {\bibfnamefont
  {Jean-Fran\c{c}ois}\ \bibnamefont {Paquet}},\ }\bibfield  {title} {\enquote
  {\bibinfo {title} {{Early-times Yang-Mills dynamics and the characterization
  of strongly interacting matter with statistical learning}},}\ }\href@noop {}
  {\  (\bibinfo {year} {2023}{\natexlab{a}})},\ \Eprint
  {http://arxiv.org/abs/2306.09619} {arXiv:2306.09619 [nucl-th]} \BibitemShut
  {NoStop}%
\bibitem [{\citenamefont {Heffernan}\ \emph
  {et~al.}(2023{\natexlab{b}})\citenamefont {Heffernan}, \citenamefont {Gale},
  \citenamefont {Jeon},\ and\ \citenamefont {Paquet}}]{Heffernan:2023utr}%
  \BibitemOpen
  \bibfield  {author} {\bibinfo {author} {\bibfnamefont {Matthew~R.}\
  \bibnamefont {Heffernan}}, \bibinfo {author} {\bibfnamefont {Charles}\
  \bibnamefont {Gale}}, \bibinfo {author} {\bibfnamefont {Sangyong}\
  \bibnamefont {Jeon}}, \ and\ \bibinfo {author} {\bibfnamefont
  {Jean-Fran\c{c}ois}\ \bibnamefont {Paquet}},\ }\bibfield  {title} {\enquote
  {\bibinfo {title} {{Bayesian quantification of strongly-interacting matter
  with color glass condensate initial conditions}},}\ }\href@noop {} {\
  (\bibinfo {year} {2023}{\natexlab{b}})},\ \Eprint
  {http://arxiv.org/abs/2302.09478} {arXiv:2302.09478 [nucl-th]} \BibitemShut
  {NoStop}%
\bibitem [{\citenamefont {Shen}\ and\ \citenamefont
  {Schenke}(2018)}]{Shen:2017bsr}%
  \BibitemOpen
  \bibfield  {author} {\bibinfo {author} {\bibfnamefont {Chun}\ \bibnamefont
  {Shen}}\ and\ \bibinfo {author} {\bibfnamefont {Bj\"orn}\ \bibnamefont
  {Schenke}},\ }\bibfield  {title} {\enquote {\bibinfo {title} {{Dynamical
  initial state model for relativistic heavy-ion collisions}},}\ }\href
  {\doibase 10.1103/PhysRevC.97.024907} {\bibfield  {journal} {\bibinfo
  {journal} {Phys. Rev. C}\ }\textbf {\bibinfo {volume} {97}},\ \bibinfo
  {pages} {024907} (\bibinfo {year} {2018})},\ \Eprint
  {http://arxiv.org/abs/1710.00881} {arXiv:1710.00881 [nucl-th]} \BibitemShut
  {NoStop}%
\bibitem [{\citenamefont {Shen}\ and\ \citenamefont
  {Schenke}(2022)}]{Shen:2022oyg}%
  \BibitemOpen
  \bibfield  {author} {\bibinfo {author} {\bibfnamefont {Chun}\ \bibnamefont
  {Shen}}\ and\ \bibinfo {author} {\bibfnamefont {Bj\"orn}\ \bibnamefont
  {Schenke}},\ }\bibfield  {title} {\enquote {\bibinfo {title} {{Longitudinal
  dynamics and particle production in relativistic nuclear collisions}},}\
  }\href {\doibase 10.1103/PhysRevC.105.064905} {\bibfield  {journal} {\bibinfo
   {journal} {Phys. Rev. C}\ }\textbf {\bibinfo {volume} {105}},\ \bibinfo
  {pages} {064905} (\bibinfo {year} {2022})},\ \Eprint
  {http://arxiv.org/abs/2203.04685} {arXiv:2203.04685 [nucl-th]} \BibitemShut
  {NoStop}%
\bibitem [{\citenamefont {Zhao}\ \emph {et~al.}(2023)\citenamefont {Zhao},
  \citenamefont {Ryu}, \citenamefont {Shen},\ and\ \citenamefont
  {Schenke}}]{Zhao:2022ugy}%
  \BibitemOpen
  \bibfield  {author} {\bibinfo {author} {\bibfnamefont {Wenbin}\ \bibnamefont
  {Zhao}}, \bibinfo {author} {\bibfnamefont {Sangwook}\ \bibnamefont {Ryu}},
  \bibinfo {author} {\bibfnamefont {Chun}\ \bibnamefont {Shen}}, \ and\
  \bibinfo {author} {\bibfnamefont {Bj\"orn}\ \bibnamefont {Schenke}},\
  }\bibfield  {title} {\enquote {\bibinfo {title} {{3D structure of anisotropic
  flow in small collision systems at energies available at the BNL Relativistic
  Heavy Ion Collider}},}\ }\href {\doibase 10.1103/PhysRevC.107.014904}
  {\bibfield  {journal} {\bibinfo  {journal} {Phys. Rev. C}\ }\textbf {\bibinfo
  {volume} {107}},\ \bibinfo {pages} {014904} (\bibinfo {year} {2023})},\
  \Eprint {http://arxiv.org/abs/2211.16376} {arXiv:2211.16376 [nucl-th]}
  \BibitemShut {NoStop}%
\bibitem [{\citenamefont {Monnai}\ \emph {et~al.}(2019)\citenamefont {Monnai},
  \citenamefont {Schenke},\ and\ \citenamefont {Shen}}]{Monnai:2019hkn}%
  \BibitemOpen
  \bibfield  {author} {\bibinfo {author} {\bibfnamefont {Akihiko}\ \bibnamefont
  {Monnai}}, \bibinfo {author} {\bibfnamefont {Bj\"orn}\ \bibnamefont
  {Schenke}}, \ and\ \bibinfo {author} {\bibfnamefont {Chun}\ \bibnamefont
  {Shen}},\ }\bibfield  {title} {\enquote {\bibinfo {title} {{Equation of state
  at finite densities for QCD matter in nuclear collisions}},}\ }\href
  {\doibase 10.1103/PhysRevC.100.024907} {\bibfield  {journal} {\bibinfo
  {journal} {Phys. Rev. C}\ }\textbf {\bibinfo {volume} {100}},\ \bibinfo
  {pages} {024907} (\bibinfo {year} {2019})},\ \Eprint
  {http://arxiv.org/abs/1902.05095} {arXiv:1902.05095 [nucl-th]} \BibitemShut
  {NoStop}%
\bibitem [{\citenamefont {Schenke}\ \emph {et~al.}(2010)\citenamefont
  {Schenke}, \citenamefont {Jeon},\ and\ \citenamefont
  {Gale}}]{Schenke:2010nt}%
  \BibitemOpen
  \bibfield  {author} {\bibinfo {author} {\bibfnamefont {Bjoern}\ \bibnamefont
  {Schenke}}, \bibinfo {author} {\bibfnamefont {Sangyong}\ \bibnamefont
  {Jeon}}, \ and\ \bibinfo {author} {\bibfnamefont {Charles}\ \bibnamefont
  {Gale}},\ }\bibfield  {title} {\enquote {\bibinfo {title} {{(3+1)D
  hydrodynamic simulation of relativistic heavy-ion collisions}},}\ }\href
  {\doibase 10.1103/PhysRevC.82.014903} {\bibfield  {journal} {\bibinfo
  {journal} {Phys. Rev. C}\ }\textbf {\bibinfo {volume} {82}},\ \bibinfo
  {pages} {014903} (\bibinfo {year} {2010})},\ \Eprint
  {http://arxiv.org/abs/1004.1408} {arXiv:1004.1408 [hep-ph]} \BibitemShut
  {NoStop}%
\bibitem [{\citenamefont {Paquet}\ \emph {et~al.}(2016)\citenamefont {Paquet},
  \citenamefont {Shen}, \citenamefont {Denicol}, \citenamefont {Luzum},
  \citenamefont {Schenke}, \citenamefont {Jeon},\ and\ \citenamefont
  {Gale}}]{Paquet:2015lta}%
  \BibitemOpen
  \bibfield  {author} {\bibinfo {author} {\bibfnamefont {Jean-Fran\c{c}ois}\
  \bibnamefont {Paquet}}, \bibinfo {author} {\bibfnamefont {Chun}\ \bibnamefont
  {Shen}}, \bibinfo {author} {\bibfnamefont {Gabriel~S.}\ \bibnamefont
  {Denicol}}, \bibinfo {author} {\bibfnamefont {Matthew}\ \bibnamefont
  {Luzum}}, \bibinfo {author} {\bibfnamefont {Bj\"orn}\ \bibnamefont
  {Schenke}}, \bibinfo {author} {\bibfnamefont {Sangyong}\ \bibnamefont
  {Jeon}}, \ and\ \bibinfo {author} {\bibfnamefont {Charles}\ \bibnamefont
  {Gale}},\ }\bibfield  {title} {\enquote {\bibinfo {title} {{Production of
  photons in relativistic heavy-ion collisions}},}\ }\href {\doibase
  10.1103/PhysRevC.93.044906} {\bibfield  {journal} {\bibinfo  {journal} {Phys.
  Rev. C}\ }\textbf {\bibinfo {volume} {93}},\ \bibinfo {pages} {044906}
  (\bibinfo {year} {2016})},\ \Eprint {http://arxiv.org/abs/1509.06738}
  {arXiv:1509.06738 [hep-ph]} \BibitemShut {NoStop}%
\bibitem [{\citenamefont {Denicol}\ \emph {et~al.}(2018)\citenamefont
  {Denicol}, \citenamefont {Gale}, \citenamefont {Jeon}, \citenamefont
  {Monnai}, \citenamefont {Schenke},\ and\ \citenamefont
  {Shen}}]{Denicol:2018wdp}%
  \BibitemOpen
  \bibfield  {author} {\bibinfo {author} {\bibfnamefont {Gabriel~S.}\
  \bibnamefont {Denicol}}, \bibinfo {author} {\bibfnamefont {Charles}\
  \bibnamefont {Gale}}, \bibinfo {author} {\bibfnamefont {Sangyong}\
  \bibnamefont {Jeon}}, \bibinfo {author} {\bibfnamefont {Akihiko}\
  \bibnamefont {Monnai}}, \bibinfo {author} {\bibfnamefont {Bj\"orn}\
  \bibnamefont {Schenke}}, \ and\ \bibinfo {author} {\bibfnamefont {Chun}\
  \bibnamefont {Shen}},\ }\bibfield  {title} {\enquote {\bibinfo {title} {{Net
  baryon diffusion in fluid dynamic simulations of relativistic heavy-ion
  collisions}},}\ }\href {\doibase 10.1103/PhysRevC.98.034916} {\bibfield
  {journal} {\bibinfo  {journal} {Phys. Rev. C}\ }\textbf {\bibinfo {volume}
  {98}},\ \bibinfo {pages} {034916} (\bibinfo {year} {2018})},\ \Eprint
  {http://arxiv.org/abs/1804.10557} {arXiv:1804.10557 [nucl-th]} \BibitemShut
  {NoStop}%
\bibitem [{\citenamefont {Bazavov}\ \emph {et~al.}(2019)\citenamefont {Bazavov}
  \emph {et~al.}}]{HotQCD:2018pds}%
  \BibitemOpen
  \bibfield  {author} {\bibinfo {author} {\bibfnamefont {A.}~\bibnamefont
  {Bazavov}} \emph {et~al.} (\bibinfo {collaboration} {HotQCD}),\ }\bibfield
  {title} {\enquote {\bibinfo {title} {{Chiral crossover in QCD at zero and
  non-zero chemical potentials}},}\ }\href {\doibase
  10.1016/j.physletb.2019.05.013} {\bibfield  {journal} {\bibinfo  {journal}
  {Phys. Lett. B}\ }\textbf {\bibinfo {volume} {795}},\ \bibinfo {pages}
  {15--21} (\bibinfo {year} {2019})},\ \Eprint
  {http://arxiv.org/abs/1812.08235} {arXiv:1812.08235 [hep-lat]} \BibitemShut
  {NoStop}%
\bibitem [{\citenamefont {Borsanyi}\ \emph {et~al.}(2020)\citenamefont
  {Borsanyi}, \citenamefont {Fodor}, \citenamefont {Guenther}, \citenamefont
  {Kara}, \citenamefont {Katz}, \citenamefont {Parotto}, \citenamefont
  {Pasztor}, \citenamefont {Ratti},\ and\ \citenamefont
  {Szabo}}]{Borsanyi:2020fev}%
  \BibitemOpen
  \bibfield  {author} {\bibinfo {author} {\bibfnamefont {Szabolcs}\
  \bibnamefont {Borsanyi}}, \bibinfo {author} {\bibfnamefont {Zoltan}\
  \bibnamefont {Fodor}}, \bibinfo {author} {\bibfnamefont {Jana~N.}\
  \bibnamefont {Guenther}}, \bibinfo {author} {\bibfnamefont {Ruben}\
  \bibnamefont {Kara}}, \bibinfo {author} {\bibfnamefont {Sandor~D.}\
  \bibnamefont {Katz}}, \bibinfo {author} {\bibfnamefont {Paolo}\ \bibnamefont
  {Parotto}}, \bibinfo {author} {\bibfnamefont {Attila}\ \bibnamefont
  {Pasztor}}, \bibinfo {author} {\bibfnamefont {Claudia}\ \bibnamefont
  {Ratti}}, \ and\ \bibinfo {author} {\bibfnamefont {Kalman~K.}\ \bibnamefont
  {Szabo}},\ }\bibfield  {title} {\enquote {\bibinfo {title} {{QCD Crossover at
  Finite Chemical Potential from Lattice Simulations}},}\ }\href {\doibase
  10.1103/PhysRevLett.125.052001} {\bibfield  {journal} {\bibinfo  {journal}
  {Phys. Rev. Lett.}\ }\textbf {\bibinfo {volume} {125}},\ \bibinfo {pages}
  {052001} (\bibinfo {year} {2020})},\ \Eprint
  {http://arxiv.org/abs/2002.02821} {arXiv:2002.02821 [hep-lat]} \BibitemShut
  {NoStop}%
\bibitem [{\citenamefont {Bass}\ \emph {et~al.}(1998)\citenamefont {Bass} \emph
  {et~al.}}]{Bass:1998ca}%
  \BibitemOpen
  \bibfield  {author} {\bibinfo {author} {\bibfnamefont {S.~A.}\ \bibnamefont
  {Bass}} \emph {et~al.},\ }\bibfield  {title} {\enquote {\bibinfo {title}
  {{Microscopic models for ultrarelativistic heavy ion collisions}},}\ }\href
  {\doibase 10.1016/S0146-6410(98)00058-1} {\bibfield  {journal} {\bibinfo
  {journal} {Prog. Part. Nucl. Phys.}\ }\textbf {\bibinfo {volume} {41}},\
  \bibinfo {pages} {255--369} (\bibinfo {year} {1998})},\ \Eprint
  {http://arxiv.org/abs/nucl-th/9803035} {arXiv:nucl-th/9803035} \BibitemShut
  {NoStop}%
\bibitem [{\citenamefont {Bleicher}\ \emph {et~al.}(1999)\citenamefont
  {Bleicher} \emph {et~al.}}]{Bleicher:1999xi}%
  \BibitemOpen
  \bibfield  {author} {\bibinfo {author} {\bibfnamefont {M.}~\bibnamefont
  {Bleicher}} \emph {et~al.},\ }\bibfield  {title} {\enquote {\bibinfo {title}
  {{Relativistic hadron hadron collisions in the ultrarelativistic quantum
  molecular dynamics model}},}\ }\href {\doibase 10.1088/0954-3899/25/9/308}
  {\bibfield  {journal} {\bibinfo  {journal} {J. Phys. G}\ }\textbf {\bibinfo
  {volume} {25}},\ \bibinfo {pages} {1859--1896} (\bibinfo {year} {1999})},\
  \Eprint {http://arxiv.org/abs/hep-ph/9909407} {arXiv:hep-ph/9909407}
  \BibitemShut {NoStop}%
\bibitem [{\citenamefont {Abelev}\ \emph {et~al.}(2009)\citenamefont {Abelev}
  \emph {et~al.}}]{STAR:2008med}%
  \BibitemOpen
  \bibfield  {author} {\bibinfo {author} {\bibfnamefont {B.~I.}\ \bibnamefont
  {Abelev}} \emph {et~al.} (\bibinfo {collaboration} {STAR}),\ }\bibfield
  {title} {\enquote {\bibinfo {title} {{Systematic Measurements of Identified
  Particle Spectra in $p p, d^+$ Au and Au+Au Collisions from STAR}},}\ }\href
  {\doibase 10.1103/PhysRevC.79.034909} {\bibfield  {journal} {\bibinfo
  {journal} {Phys. Rev. C}\ }\textbf {\bibinfo {volume} {79}},\ \bibinfo
  {pages} {034909} (\bibinfo {year} {2009})},\ \Eprint
  {http://arxiv.org/abs/0808.2041} {arXiv:0808.2041 [nucl-ex]} \BibitemShut
  {NoStop}%
\bibitem [{\citenamefont {Back}\ \emph {et~al.}(2006)\citenamefont {Back} \emph
  {et~al.}}]{PHOBOS:2005zhy}%
  \BibitemOpen
  \bibfield  {author} {\bibinfo {author} {\bibfnamefont {B.~B.}\ \bibnamefont
  {Back}} \emph {et~al.} (\bibinfo {collaboration} {PHOBOS}),\ }\bibfield
  {title} {\enquote {\bibinfo {title} {{Charged-particle pseudorapidity
  distributions in Au+Au collisions at $s(NN) ^{1/2}$ = 62.4-GeV}},}\ }\href
  {\doibase 10.1103/PhysRevC.74.021901} {\bibfield  {journal} {\bibinfo
  {journal} {Phys. Rev. C}\ }\textbf {\bibinfo {volume} {74}},\ \bibinfo
  {pages} {021901} (\bibinfo {year} {2006})},\ \Eprint
  {http://arxiv.org/abs/nucl-ex/0509034} {arXiv:nucl-ex/0509034} \BibitemShut
  {NoStop}%
\bibitem [{\citenamefont {Alver}\ \emph {et~al.}(2007)\citenamefont {Alver}
  \emph {et~al.}}]{PHOBOS:2006dbo}%
  \BibitemOpen
  \bibfield  {author} {\bibinfo {author} {\bibfnamefont {B.}~\bibnamefont
  {Alver}} \emph {et~al.} (\bibinfo {collaboration} {PHOBOS}),\ }\bibfield
  {title} {\enquote {\bibinfo {title} {{System size, energy, pseudorapidity,
  and centrality dependence of elliptic flow}},}\ }\href {\doibase
  10.1103/PhysRevLett.98.242302} {\bibfield  {journal} {\bibinfo  {journal}
  {Phys. Rev. Lett.}\ }\textbf {\bibinfo {volume} {98}},\ \bibinfo {pages}
  {242302} (\bibinfo {year} {2007})},\ \Eprint
  {http://arxiv.org/abs/nucl-ex/0610037} {arXiv:nucl-ex/0610037} \BibitemShut
  {NoStop}%
\bibitem [{\citenamefont {Adamczyk}\ \emph {et~al.}(2018)\citenamefont
  {Adamczyk} \emph {et~al.}}]{STAR:2017idk}%
  \BibitemOpen
  \bibfield  {author} {\bibinfo {author} {\bibfnamefont {L.}~\bibnamefont
  {Adamczyk}} \emph {et~al.} (\bibinfo {collaboration} {STAR}),\ }\bibfield
  {title} {\enquote {\bibinfo {title} {{Harmonic decomposition of
  three-particle azimuthal correlations at energies available at the BNL
  Relativistic Heavy Ion Collider}},}\ }\href {\doibase
  10.1103/PhysRevC.98.034918} {\bibfield  {journal} {\bibinfo  {journal} {Phys.
  Rev. C}\ }\textbf {\bibinfo {volume} {98}},\ \bibinfo {pages} {034918}
  (\bibinfo {year} {2018})},\ \Eprint {http://arxiv.org/abs/1701.06496}
  {arXiv:1701.06496 [nucl-ex]} \BibitemShut {NoStop}%
\bibitem [{\citenamefont {Adamczyk}\ \emph {et~al.}(2016)\citenamefont
  {Adamczyk} \emph {et~al.}}]{STAR:2016vqt}%
  \BibitemOpen
  \bibfield  {author} {\bibinfo {author} {\bibfnamefont {L.}~\bibnamefont
  {Adamczyk}} \emph {et~al.} (\bibinfo {collaboration} {STAR}),\ }\bibfield
  {title} {\enquote {\bibinfo {title} {{Beam Energy Dependence of the Third
  Harmonic of Azimuthal Correlations in Au+Au Collisions at RHIC}},}\ }\href
  {\doibase 10.1103/PhysRevLett.116.112302} {\bibfield  {journal} {\bibinfo
  {journal} {Phys. Rev. Lett.}\ }\textbf {\bibinfo {volume} {116}},\ \bibinfo
  {pages} {112302} (\bibinfo {year} {2016})},\ \Eprint
  {http://arxiv.org/abs/1601.01999} {arXiv:1601.01999 [nucl-ex]} \BibitemShut
  {NoStop}%
\bibitem [{\citenamefont {Adamczyk}\ \emph {et~al.}(2017)\citenamefont
  {Adamczyk} \emph {et~al.}}]{STAR:2017sal}%
  \BibitemOpen
  \bibfield  {author} {\bibinfo {author} {\bibfnamefont {L.}~\bibnamefont
  {Adamczyk}} \emph {et~al.} (\bibinfo {collaboration} {STAR}),\ }\bibfield
  {title} {\enquote {\bibinfo {title} {{Bulk Properties of the Medium Produced
  in Relativistic Heavy-Ion Collisions from the Beam Energy Scan Program}},}\
  }\href {\doibase 10.1103/PhysRevC.96.044904} {\bibfield  {journal} {\bibinfo
  {journal} {Phys. Rev. C}\ }\textbf {\bibinfo {volume} {96}},\ \bibinfo
  {pages} {044904} (\bibinfo {year} {2017})},\ \Eprint
  {http://arxiv.org/abs/1701.07065} {arXiv:1701.07065 [nucl-ex]} \BibitemShut
  {NoStop}%
\bibitem [{\citenamefont {Cleymans}\ \emph {et~al.}(2006)\citenamefont
  {Cleymans}, \citenamefont {Oeschler}, \citenamefont {Redlich},\ and\
  \citenamefont {Wheaton}}]{Cleymans:2005xv}%
  \BibitemOpen
  \bibfield  {author} {\bibinfo {author} {\bibfnamefont {J.}~\bibnamefont
  {Cleymans}}, \bibinfo {author} {\bibfnamefont {H.}~\bibnamefont {Oeschler}},
  \bibinfo {author} {\bibfnamefont {K.}~\bibnamefont {Redlich}}, \ and\
  \bibinfo {author} {\bibfnamefont {S.}~\bibnamefont {Wheaton}},\ }\bibfield
  {title} {\enquote {\bibinfo {title} {{Comparison of chemical freeze-out
  criteria in heavy-ion collisions}},}\ }\href {\doibase
  10.1103/PhysRevC.73.034905} {\bibfield  {journal} {\bibinfo  {journal} {Phys.
  Rev. C}\ }\textbf {\bibinfo {volume} {73}},\ \bibinfo {pages} {034905}
  (\bibinfo {year} {2006})},\ \Eprint {http://arxiv.org/abs/hep-ph/0511094}
  {arXiv:hep-ph/0511094} \BibitemShut {NoStop}%
\bibitem [{\citenamefont {Andronic}\ \emph {et~al.}(2010)\citenamefont
  {Andronic}, \citenamefont {Braun-Munzinger},\ and\ \citenamefont
  {Stachel}}]{Andronic:2009jd}%
  \BibitemOpen
  \bibfield  {author} {\bibinfo {author} {\bibfnamefont {A.}~\bibnamefont
  {Andronic}}, \bibinfo {author} {\bibfnamefont {P.}~\bibnamefont
  {Braun-Munzinger}}, \ and\ \bibinfo {author} {\bibfnamefont {J.}~\bibnamefont
  {Stachel}},\ }\bibfield  {title} {\enquote {\bibinfo {title} {{The Horn, the
  hadron mass spectrum and the QCD phase diagram: The Statistical model of
  hadron production in central nucleus-nucleus collisions}},}\ }\href {\doibase
  10.1016/j.nuclphysa.2009.12.048} {\bibfield  {journal} {\bibinfo  {journal}
  {Nucl. Phys. A}\ }\textbf {\bibinfo {volume} {834}},\ \bibinfo {pages}
  {237C--240C} (\bibinfo {year} {2010})},\ \Eprint
  {http://arxiv.org/abs/0911.4931} {arXiv:0911.4931 [nucl-th]} \BibitemShut
  {NoStop}%
\bibitem [{\citenamefont {V.~Roshan~Joseph}\ and\ \citenamefont
  {Ba}(2020)}]{MaxProLHD1}%
  \BibitemOpen
  \bibfield  {author} {\bibinfo {author} {\bibfnamefont {Evren~Gul}\
  \bibnamefont {V.~Roshan~Joseph}}\ and\ \bibinfo {author} {\bibfnamefont
  {Shan}\ \bibnamefont {Ba}},\ }\bibfield  {title} {\enquote {\bibinfo {title}
  {Designing computer experiments with multiple types of factors: The maxpro
  approach},}\ }\href {\doibase 10.1080/00224065.2019.1611351} {\bibfield
  {journal} {\bibinfo  {journal} {Journal of Quality Technology}\ }\textbf
  {\bibinfo {volume} {52}},\ \bibinfo {pages} {343--354} (\bibinfo {year}
  {2020})},\ \Eprint
  {http://arxiv.org/abs/https://doi.org/10.1080/00224065.2019.1611351}
  {https://doi.org/10.1080/00224065.2019.1611351} \BibitemShut {NoStop}%
\bibitem [{\citenamefont {Joseph}\ \emph {et~al.}(2015)\citenamefont {Joseph},
  \citenamefont {Gul},\ and\ \citenamefont {Ba}}]{MaxProLHD2}%
  \BibitemOpen
  \bibfield  {author} {\bibinfo {author} {\bibfnamefont {V.~Roshan}\
  \bibnamefont {Joseph}}, \bibinfo {author} {\bibfnamefont {Evren}\
  \bibnamefont {Gul}}, \ and\ \bibinfo {author} {\bibfnamefont {Shan}\
  \bibnamefont {Ba}},\ }\bibfield  {title} {\enquote {\bibinfo {title}
  {{Maximum projection designs for computer experiments}},}\ }\href {\doibase
  10.1093/biomet/asv002} {\bibfield  {journal} {\bibinfo  {journal}
  {Biometrika}\ }\textbf {\bibinfo {volume} {102}},\ \bibinfo {pages}
  {371--380} (\bibinfo {year} {2015})},\ \Eprint
  {http://arxiv.org/abs/https://academic.oup.com/biomet/article-pdf/102/2/371/9642501/asv002.pdf}
  {https://academic.oup.com/biomet/article-pdf/102/2/371/9642501/asv002.pdf}
  \BibitemShut {NoStop}%
\bibitem [{str()}]{streamlit}%
  \BibitemOpen
  \href@noop {} {}\bibinfo {note} {The interactive web page is powered by
  Streamlit,
  \url{https://3dglauberappapp-bbbfxt8w75lbvnkcxreahl.streamlit.app}.}\BibitemShut
  {Stop}%
\bibitem [{\citenamefont {M\"antysaari}\ \emph {et~al.}(2022)\citenamefont
  {M\"antysaari}, \citenamefont {Schenke}, \citenamefont {Shen},\ and\
  \citenamefont {Zhao}}]{Mantysaari:2022ffw}%
  \BibitemOpen
  \bibfield  {author} {\bibinfo {author} {\bibfnamefont {Heikki}\ \bibnamefont
  {M\"antysaari}}, \bibinfo {author} {\bibfnamefont {Bj\"orn}\ \bibnamefont
  {Schenke}}, \bibinfo {author} {\bibfnamefont {Chun}\ \bibnamefont {Shen}}, \
  and\ \bibinfo {author} {\bibfnamefont {Wenbin}\ \bibnamefont {Zhao}},\
  }\bibfield  {title} {\enquote {\bibinfo {title} {{Bayesian inference of the
  fluctuating proton shape}},}\ }\href {\doibase
  10.1016/j.physletb.2022.137348} {\bibfield  {journal} {\bibinfo  {journal}
  {Phys. Lett. B}\ }\textbf {\bibinfo {volume} {833}},\ \bibinfo {pages}
  {137348} (\bibinfo {year} {2022})},\ \Eprint
  {http://arxiv.org/abs/2202.01998} {arXiv:2202.01998 [hep-ph]} \BibitemShut
  {NoStop}%
\bibitem [{\citenamefont {Videb\ae{}k}\ and\ \citenamefont
  {Hansen}(1995)}]{PhysRevC.52.2684}%
  \BibitemOpen
  \bibfield  {author} {\bibinfo {author} {\bibfnamefont {F.}~\bibnamefont
  {Videb\ae{}k}}\ and\ \bibinfo {author} {\bibfnamefont {Ole}\ \bibnamefont
  {Hansen}},\ }\bibfield  {title} {\enquote {\bibinfo {title} {Baryon rapidity
  loss and midrapidity stacking in high energy nucleus-nucleus collisions},}\
  }\href {\doibase 10.1103/PhysRevC.52.2684} {\bibfield  {journal} {\bibinfo
  {journal} {Phys. Rev. C}\ }\textbf {\bibinfo {volume} {52}},\ \bibinfo
  {pages} {2684--2693} (\bibinfo {year} {1995})}\BibitemShut {NoStop}%
\bibitem [{\citenamefont {Arsene}\ \emph {et~al.}(2009)\citenamefont {Arsene}
  \emph {et~al.}}]{BRAHMS:2009wlg}%
  \BibitemOpen
  \bibfield  {author} {\bibinfo {author} {\bibfnamefont {I.~C.}\ \bibnamefont
  {Arsene}} \emph {et~al.} (\bibinfo {collaboration} {BRAHMS}),\ }\bibfield
  {title} {\enquote {\bibinfo {title} {{Nuclear stopping and rapidity loss in
  Au+Au collisions at s(NN)**(1/2) = 62.4-GeV}},}\ }\href {\doibase
  10.1016/j.physletb.2009.05.049} {\bibfield  {journal} {\bibinfo  {journal}
  {Phys. Lett. B}\ }\textbf {\bibinfo {volume} {677}},\ \bibinfo {pages}
  {267--271} (\bibinfo {year} {2009})},\ \Eprint
  {http://arxiv.org/abs/0901.0872} {arXiv:0901.0872 [nucl-ex]} \BibitemShut
  {NoStop}%
\bibitem [{\citenamefont {Shen}\ and\ \citenamefont
  {Schenke}(2019)}]{Shen:2018pty}%
  \BibitemOpen
  \bibfield  {author} {\bibinfo {author} {\bibfnamefont {Chun}\ \bibnamefont
  {Shen}}\ and\ \bibinfo {author} {\bibfnamefont {Bj\"orn}\ \bibnamefont
  {Schenke}},\ }\bibfield  {title} {\enquote {\bibinfo {title} {{Dynamical
  initialization and hydrodynamic modeling of relativistic heavy-ion
  collisions}},}\ }\href {\doibase 10.1016/j.nuclphysa.2018.08.007} {\bibfield
  {journal} {\bibinfo  {journal} {Nucl. Phys. A}\ }\textbf {\bibinfo {volume}
  {982}},\ \bibinfo {pages} {411--414} (\bibinfo {year} {2019})},\ \Eprint
  {http://arxiv.org/abs/1807.05141} {arXiv:1807.05141 [nucl-th]} \BibitemShut
  {NoStop}%
\bibitem [{\citenamefont {Denicol}\ \emph {et~al.}(2016)\citenamefont
  {Denicol}, \citenamefont {Monnai},\ and\ \citenamefont
  {Schenke}}]{Denicol:2015nhu}%
  \BibitemOpen
  \bibfield  {author} {\bibinfo {author} {\bibfnamefont {Gabriel}\ \bibnamefont
  {Denicol}}, \bibinfo {author} {\bibfnamefont {Akihiko}\ \bibnamefont
  {Monnai}}, \ and\ \bibinfo {author} {\bibfnamefont {Bjoern}\ \bibnamefont
  {Schenke}},\ }\bibfield  {title} {\enquote {\bibinfo {title} {{Moving forward
  to constrain the shear viscosity of QCD matter}},}\ }\href {\doibase
  10.1103/PhysRevLett.116.212301} {\bibfield  {journal} {\bibinfo  {journal}
  {Phys. Rev. Lett.}\ }\textbf {\bibinfo {volume} {116}},\ \bibinfo {pages}
  {212301} (\bibinfo {year} {2016})},\ \Eprint
  {http://arxiv.org/abs/1512.01538} {arXiv:1512.01538 [nucl-th]} \BibitemShut
  {NoStop}%
\bibitem [{\citenamefont {Hoyos}\ \emph {et~al.}(2020)\citenamefont {Hoyos},
  \citenamefont {Jokela}, \citenamefont {Jarvinen}, \citenamefont {Subils},
  \citenamefont {Tarrio},\ and\ \citenamefont {Vuorinen}}]{Hoyos:2020hmq}%
  \BibitemOpen
  \bibfield  {author} {\bibinfo {author} {\bibfnamefont {Carlos}\ \bibnamefont
  {Hoyos}}, \bibinfo {author} {\bibfnamefont {Niko}\ \bibnamefont {Jokela}},
  \bibinfo {author} {\bibfnamefont {Matti}\ \bibnamefont {Jarvinen}}, \bibinfo
  {author} {\bibfnamefont {Javier~G.}\ \bibnamefont {Subils}}, \bibinfo
  {author} {\bibfnamefont {Javier}\ \bibnamefont {Tarrio}}, \ and\ \bibinfo
  {author} {\bibfnamefont {Aleksi}\ \bibnamefont {Vuorinen}},\ }\bibfield
  {title} {\enquote {\bibinfo {title} {{Transport in strongly coupled quark
  matter}},}\ }\href {\doibase 10.1103/PhysRevLett.125.241601} {\bibfield
  {journal} {\bibinfo  {journal} {Phys. Rev. Lett.}\ }\textbf {\bibinfo
  {volume} {125}},\ \bibinfo {pages} {241601} (\bibinfo {year} {2020})},\
  \Eprint {http://arxiv.org/abs/2005.14205} {arXiv:2005.14205 [hep-th]}
  \BibitemShut {NoStop}%
\bibitem [{\citenamefont {Soloveva}\ \emph {et~al.}(2021)\citenamefont
  {Soloveva}, \citenamefont {Fuseau}, \citenamefont {Aichelin},\ and\
  \citenamefont {Bratkovskaya}}]{Soloveva:2020hpr}%
  \BibitemOpen
  \bibfield  {author} {\bibinfo {author} {\bibfnamefont {Olga}\ \bibnamefont
  {Soloveva}}, \bibinfo {author} {\bibfnamefont {David}\ \bibnamefont
  {Fuseau}}, \bibinfo {author} {\bibfnamefont {J\"org}\ \bibnamefont
  {Aichelin}}, \ and\ \bibinfo {author} {\bibfnamefont {Elena}\ \bibnamefont
  {Bratkovskaya}},\ }\bibfield  {title} {\enquote {\bibinfo {title} {{Shear
  viscosity and electric conductivity of a hot and dense QGP with a chiral
  phase transition}},}\ }\href {\doibase 10.1103/PhysRevC.103.054901}
  {\bibfield  {journal} {\bibinfo  {journal} {Phys. Rev. C}\ }\textbf {\bibinfo
  {volume} {103}},\ \bibinfo {pages} {054901} (\bibinfo {year} {2021})},\
  \Eprint {http://arxiv.org/abs/2011.03505} {arXiv:2011.03505 [nucl-th]}
  \BibitemShut {NoStop}%
\bibitem [{\citenamefont {McLaughlin}\ \emph {et~al.}(2022)\citenamefont
  {McLaughlin}, \citenamefont {Rose}, \citenamefont {Dore}, \citenamefont
  {Parotto}, \citenamefont {Ratti},\ and\ \citenamefont
  {Noronha-Hostler}}]{McLaughlin:2021dph}%
  \BibitemOpen
  \bibfield  {author} {\bibinfo {author} {\bibfnamefont {Emma}\ \bibnamefont
  {McLaughlin}}, \bibinfo {author} {\bibfnamefont {Jacob}\ \bibnamefont
  {Rose}}, \bibinfo {author} {\bibfnamefont {Travis}\ \bibnamefont {Dore}},
  \bibinfo {author} {\bibfnamefont {Paolo}\ \bibnamefont {Parotto}}, \bibinfo
  {author} {\bibfnamefont {Claudia}\ \bibnamefont {Ratti}}, \ and\ \bibinfo
  {author} {\bibfnamefont {Jacquelyn}\ \bibnamefont {Noronha-Hostler}},\
  }\bibfield  {title} {\enquote {\bibinfo {title} {{Building a testable shear
  viscosity across the QCD phase diagram}},}\ }\href {\doibase
  10.1103/PhysRevC.105.024903} {\bibfield  {journal} {\bibinfo  {journal}
  {Phys. Rev. C}\ }\textbf {\bibinfo {volume} {105}},\ \bibinfo {pages}
  {024903} (\bibinfo {year} {2022})},\ \Eprint
  {http://arxiv.org/abs/2103.02090} {arXiv:2103.02090 [nucl-th]} \BibitemShut
  {NoStop}%
\bibitem [{\citenamefont {Denicol}\ \emph {et~al.}(2013)\citenamefont
  {Denicol}, \citenamefont {Gale}, \citenamefont {Jeon},\ and\ \citenamefont
  {Noronha}}]{Denicol:2013nua}%
  \BibitemOpen
  \bibfield  {author} {\bibinfo {author} {\bibfnamefont {Gabriel~S.}\
  \bibnamefont {Denicol}}, \bibinfo {author} {\bibfnamefont {Charles}\
  \bibnamefont {Gale}}, \bibinfo {author} {\bibfnamefont {Sangyong}\
  \bibnamefont {Jeon}}, \ and\ \bibinfo {author} {\bibfnamefont {Jorge}\
  \bibnamefont {Noronha}},\ }\bibfield  {title} {\enquote {\bibinfo {title}
  {{Fluid behavior of a baryon-rich hadron resonance gas}},}\ }\href {\doibase
  10.1103/PhysRevC.88.064901} {\bibfield  {journal} {\bibinfo  {journal} {Phys.
  Rev. C}\ }\textbf {\bibinfo {volume} {88}},\ \bibinfo {pages} {064901}
  (\bibinfo {year} {2013})},\ \Eprint {http://arxiv.org/abs/1308.1923}
  {arXiv:1308.1923 [nucl-th]} \BibitemShut {NoStop}%
\bibitem [{\citenamefont {Grefa}\ \emph {et~al.}(2022)\citenamefont {Grefa},
  \citenamefont {Hippert}, \citenamefont {Noronha}, \citenamefont
  {Noronha-Hostler}, \citenamefont {Portillo}, \citenamefont {Ratti},\ and\
  \citenamefont {Rougemont}}]{Grefa:2022sav}%
  \BibitemOpen
  \bibfield  {author} {\bibinfo {author} {\bibfnamefont {Joaquin}\ \bibnamefont
  {Grefa}}, \bibinfo {author} {\bibfnamefont {Mauricio}\ \bibnamefont
  {Hippert}}, \bibinfo {author} {\bibfnamefont {Jorge}\ \bibnamefont
  {Noronha}}, \bibinfo {author} {\bibfnamefont {Jacquelyn}\ \bibnamefont
  {Noronha-Hostler}}, \bibinfo {author} {\bibfnamefont {Israel}\ \bibnamefont
  {Portillo}}, \bibinfo {author} {\bibfnamefont {Claudia}\ \bibnamefont
  {Ratti}}, \ and\ \bibinfo {author} {\bibfnamefont {Romulo}\ \bibnamefont
  {Rougemont}},\ }\bibfield  {title} {\enquote {\bibinfo {title} {{Transport
  coefficients of the quark-gluon plasma at the critical point and across the
  first-order line}},}\ }\href {\doibase 10.1103/PhysRevD.106.034024}
  {\bibfield  {journal} {\bibinfo  {journal} {Phys. Rev. D}\ }\textbf {\bibinfo
  {volume} {106}},\ \bibinfo {pages} {034024} (\bibinfo {year} {2022})},\
  \Eprint {http://arxiv.org/abs/2203.00139} {arXiv:2203.00139 [nucl-th]}
  \BibitemShut {NoStop}%
\bibitem [{\citenamefont {Adamczyk}\ \emph {et~al.}(2015)\citenamefont
  {Adamczyk} \emph {et~al.}}]{STAR:2014shf}%
  \BibitemOpen
  \bibfield  {author} {\bibinfo {author} {\bibfnamefont {L.}~\bibnamefont
  {Adamczyk}} \emph {et~al.} (\bibinfo {collaboration} {STAR}),\ }\bibfield
  {title} {\enquote {\bibinfo {title} {{Beam-energy-dependent two-pion
  interferometry and the freeze-out eccentricity of pions measured in heavy ion
  collisions at the STAR detector}},}\ }\href {\doibase
  10.1103/PhysRevC.92.014904} {\bibfield  {journal} {\bibinfo  {journal} {Phys.
  Rev. C}\ }\textbf {\bibinfo {volume} {92}},\ \bibinfo {pages} {014904}
  (\bibinfo {year} {2015})},\ \Eprint {http://arxiv.org/abs/1403.4972}
  {arXiv:1403.4972 [nucl-ex]} \BibitemShut {NoStop}%
\bibitem [{\citenamefont {Lacey}(2015)}]{Lacey:2014wqa}%
  \BibitemOpen
  \bibfield  {author} {\bibinfo {author} {\bibfnamefont {Roy~A.}\ \bibnamefont
  {Lacey}},\ }\bibfield  {title} {\enquote {\bibinfo {title} {{Indications for
  a Critical End Point in the Phase Diagram for Hot and Dense Nuclear
  Matter}},}\ }\href {\doibase 10.1103/PhysRevLett.114.142301} {\bibfield
  {journal} {\bibinfo  {journal} {Phys. Rev. Lett.}\ }\textbf {\bibinfo
  {volume} {114}},\ \bibinfo {pages} {142301} (\bibinfo {year} {2015})},\
  \Eprint {http://arxiv.org/abs/1411.7931} {arXiv:1411.7931 [nucl-ex]}
  \BibitemShut {NoStop}%
\bibitem [{\citenamefont {Pordes}\ \emph {et~al.}(2007)\citenamefont {Pordes}
  \emph {et~al.}}]{Pordes:2007zzb}%
  \BibitemOpen
  \bibfield  {author} {\bibinfo {author} {\bibfnamefont {Ruth}\ \bibnamefont
  {Pordes}} \emph {et~al.},\ }\bibfield  {title} {\enquote {\bibinfo {title}
  {{The Open Science Grid}},}\ }\href {\doibase 10.1088/1742-6596/78/1/012057}
  {\bibfield  {journal} {\bibinfo  {journal} {J. Phys. Conf. Ser.}\ }\textbf
  {\bibinfo {volume} {78}},\ \bibinfo {pages} {012057} (\bibinfo {year}
  {2007})}\BibitemShut {NoStop}%
\bibitem [{\citenamefont {Sfiligoi}\ \emph {et~al.}(2009)\citenamefont
  {Sfiligoi}, \citenamefont {Bradley}, \citenamefont {Holzman}, \citenamefont
  {Mhashilkar}, \citenamefont {Padhi},\ and\ \citenamefont
  {Wurthwrin}}]{Sfiligoi:2009cct}%
  \BibitemOpen
  \bibfield  {author} {\bibinfo {author} {\bibfnamefont {Igor}\ \bibnamefont
  {Sfiligoi}}, \bibinfo {author} {\bibfnamefont {Daniel~C.}\ \bibnamefont
  {Bradley}}, \bibinfo {author} {\bibfnamefont {Burt}\ \bibnamefont {Holzman}},
  \bibinfo {author} {\bibfnamefont {Parag}\ \bibnamefont {Mhashilkar}},
  \bibinfo {author} {\bibfnamefont {Sanjay}\ \bibnamefont {Padhi}}, \ and\
  \bibinfo {author} {\bibfnamefont {Frank}\ \bibnamefont {Wurthwrin}},\
  }\bibfield  {title} {\enquote {\bibinfo {title} {{The pilot way to Grid
  resources using glideinWMS}},}\ }\href {\doibase 10.1109/CSIE.2009.950}
  {\bibfield  {journal} {\bibinfo  {journal} {WRI World Congress}\ }\textbf
  {\bibinfo {volume} {2}},\ \bibinfo {pages} {428--432} (\bibinfo {year}
  {2009})}\BibitemShut {NoStop}%
\bibitem [{Goo()}]{GoogleDrive}%
  \BibitemOpen
  \href@noop {} {}\bibinfo {note} {The model training data can be download from
  \url{https://drive.google.com/drive/folders/1hqZVZdVKmJxMprwEyP3dxBI82Y35taP7?usp=share_link}.}\BibitemShut
  {Stop}%
\bibitem [{iEB()}]{iEBEMUSIC}%
  \BibitemOpen
  \href@noop {} {}\bibinfo {note} {The iEBE-MUSIC is a general-purpose
  numerical framework to simulate dynamical evolution of relativistic heavy-ion
  collisions event-by-event. This work uses the commit version
  \texttt{6fd7e98}, which can be downloaded from
  \url{https://github.com/chunshen1987/iEBE-MUSIC}.}\BibitemShut {Stop}%
\bibitem [{sin()}]{singularity}%
  \BibitemOpen
  \href@noop {} {}\bibinfo {note} {This work used the docker/singularity image
  with the tag \texttt{bayesian2023} to generate training data,
  \url{https://hub.docker.com/r/chunshen1987/iebe-music/tags}}\BibitemShut
  {NoStop}%
\end{thebibliography}%


\newpage
\clearpage
\setcounter{figure}{0}
\renewcommand{\thefigure}{S\arabic{figure}}
\renewcommand{\theHfigure}{S\arabic{figure}}

\widetext
\begin{center}
\textbf{\large Supplemental Material}
\end{center}

\begin{figure}[h!]
    \centering
    \includegraphics[width=0.48\linewidth]{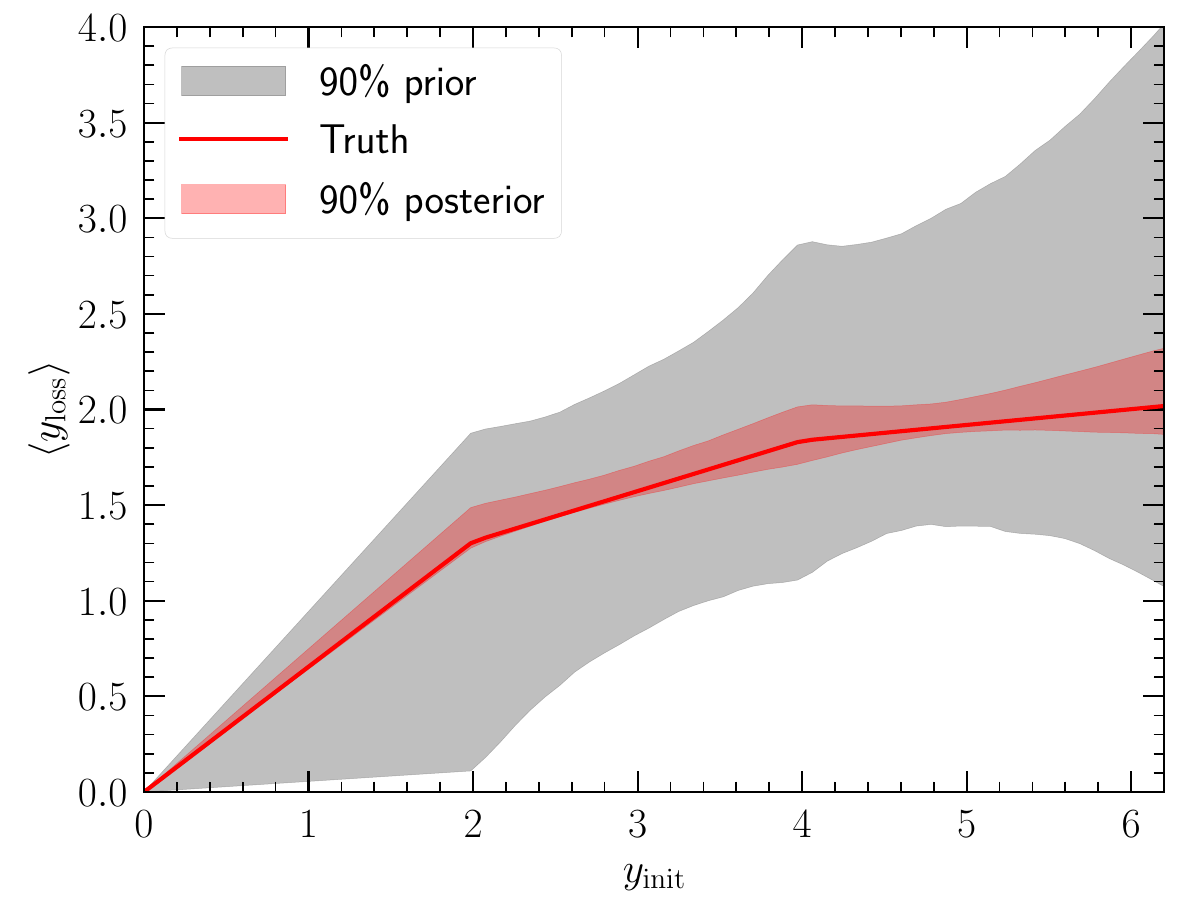}
    \includegraphics[width=0.48\linewidth]{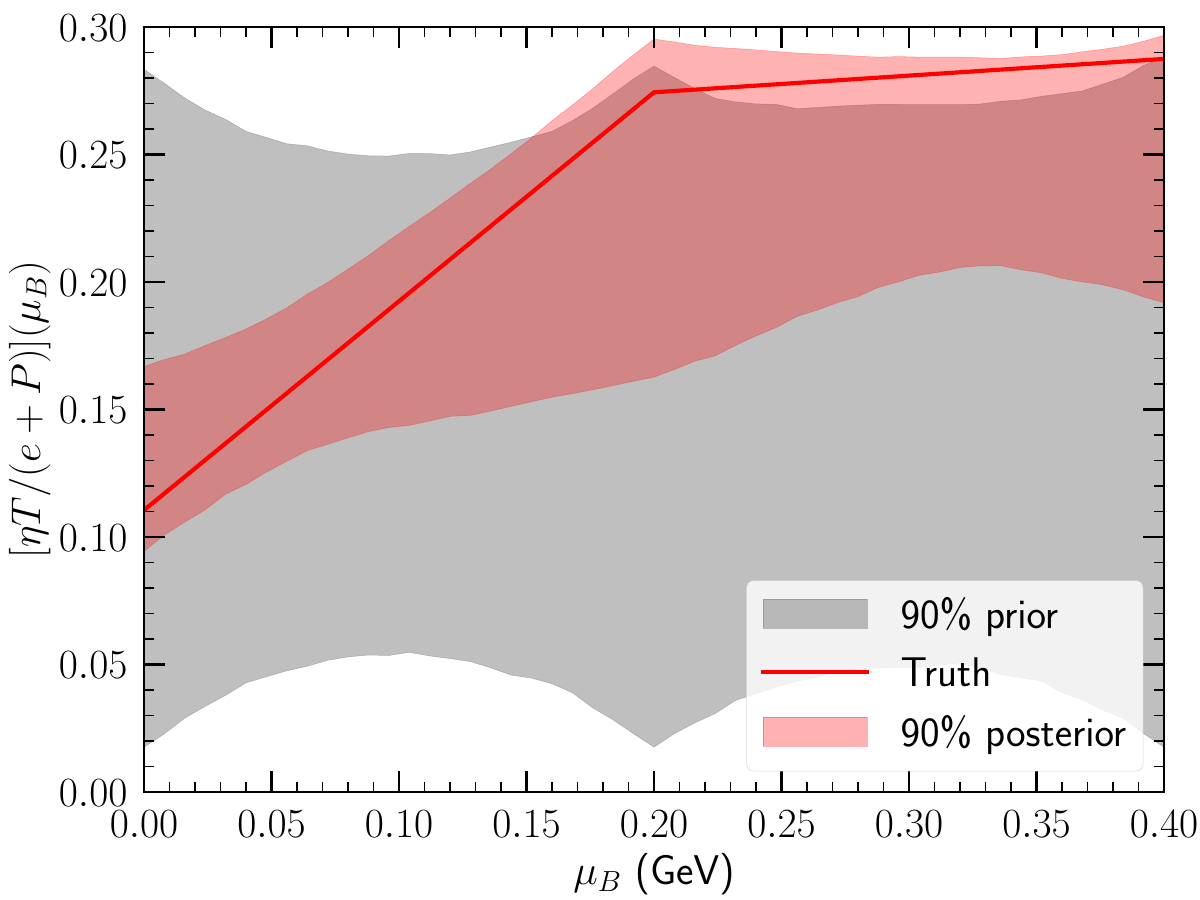}
    \includegraphics[width=0.48\linewidth]{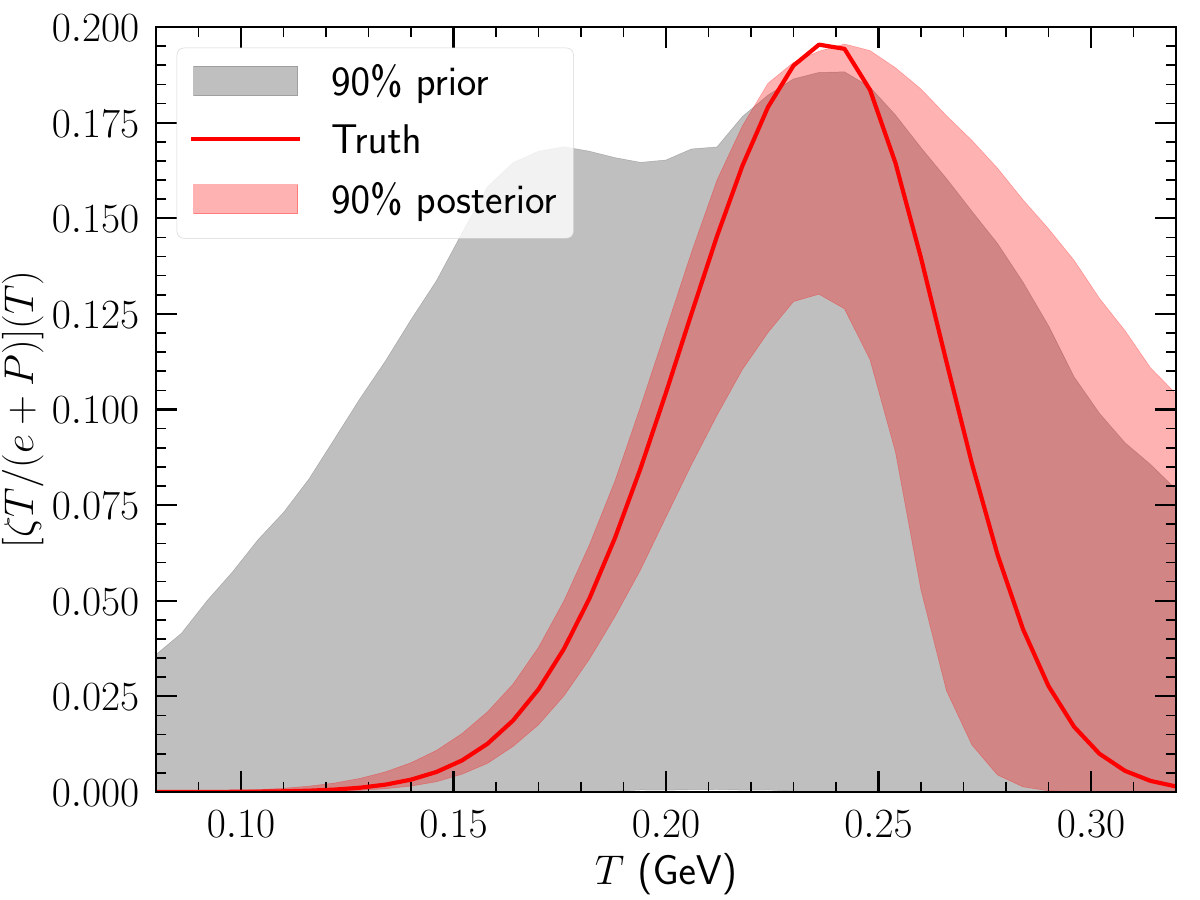}
    \caption{(Color online) Posterior distributions for initial-state average rapidity loss $\langle y_\mathrm{loss} \rangle$ and QGP specific shear and bulk viscosities compared to the true result in a closure test. The bands indicate the 90\% confidence intervals.}
    \label{fig:ClosureTest}
\end{figure}

Figure~\ref{fig:ClosureTest} shows the closure test of our Bayesian analysis performed on one of the validation parameter points. We find good agreement between the true values and the 90\% posterior regions. This closure test not only validates the functionality of our Bayesian inference analysis but also confirms the effectiveness of the selected experimental observables, that particle yields, mean transverse momenta, and flow anisotropies can provide strong constraints on the theoretical model parameters we are interested in.

All the software and data produced from this work are open source. The produced model training data can be found online~\cite{GoogleDrive}. We used the iEBE-MUSIC framework to perform all numerical simulations on the Open Science Grid (OSG). The version of the numerical framework is fixed~\cite{iEBEMUSIC}. A tagged singularity image of the software can be downloaded from the Docker Hub website~\cite{singularity}. 

\end{document}